\documentclass[useAMS,usenatbib]{mn2e}

\usepackage{times}
\usepackage{BibDef}
\usepackage{hyperref}
\usepackage{amssymb}
\usepackage{float}
\usepackage{graphicx}

\title[Early Build-up of the Hubble Sequence]{The Argo Simulation: II. The Early Build-up of the Hubble Sequence}

\author[D. Fiacconi et al.]{Davide Fiacconi$^{1,2,}$\thanks{E-mail: fiacconi@physik.uzh.ch}, 
Robert Feldmann$^{3}$\thanks{Hubble Fellow}
and Lucio Mayer$^{1,2}$\\
$^{1}$Center for Theoretical Astrophysics and Cosmology, Institute for Computational Science, University of Zurich, Winterthurerstrasse 190, CH-8057 Z\"{u}rich, Switzerland\\
$^{2}$Physik-Institut, University of Zurich, Winterthurerstrasse 190, CH-8057 Z\"{u}rich, Switzerland\\
$^{3}$Department of Astronomy, University of California, Berkeley, CA 94720-3411, USA
}

\begin{document}

\date{\today}

\pagerange{\pageref{firstpage}--\pageref{lastpage}} \pubyear{2014}

\maketitle

\label{firstpage}

%%%%%%%%%%%%%%%%%%%%%%%%%%%%%%%%%%%%%%%%%%%%%%%
%%%%%%%%%%%%%%%%%%%%%%%%%%%%%%%%%%%%%%%%%%%%%%%
%%%%%%%%%%%%%%%%%%%%%%%%%%%%%%%%%%%%%%%%%%%%%%%

\begin{abstract}
The Hubble sequence is a common classification scheme for the structure of galaxies.
Despite the tremendous usefulness of this diagnostic, we still do not fully understand when, where, and how this morphological ordering was put in place.
Here, we investigate the morphological evolution of a sample of 22 high redshift ($z\geq3$) galaxies extracted from the Argo simulation. 
Argo is a cosmological zoom-in simulation of a group-sized halo and its environment.
It adopts the same high resolution ($\sim10^4$ M$_\odot$, $\sim100$ pc) and sub-grid physical model that was used in the Eris simulation but probes a sub-volume almost ten times bigger with as many as 45 million gas and star particles in the zoom-in region.
Argo follows the early assembly of galaxies with a broad range of stellar masses ($\log M_{\star}/{\rm M}_{\odot}\sim8-11$  at $z\simeq3$), while resolving properly their structural properties.
We recover a diversity of morphologies, including late-type/irregular disc galaxies with flat rotation curves, spheroid dominated early-type discs, and a massive elliptical galaxy, already established at $z\sim3$.
We identify major mergers as the main trigger for the formation of bulges and the steepening of the circular velocity curves.
Minor mergers and non-axisymmetric perturbations (stellar bars) drive the bulge growth in some cases.
The specific angular momenta of the simulated disc components fairly match the values inferred from nearby galaxies of similar $M_{\star}$ once the expected redshift evolution of disc sizes is accounted for.
We conclude that morphological transformations of high redshift galaxies of intermediate mass are likely triggered by processes similar to those at low redshift and result in an early build-up of the Hubble sequence.
\end{abstract}


%%%%%%%%%%%%%%%%%%%%%%%%%%%%%%%%%%%%%%%%%%%%%%%
%%%%%%%%%%%%%%%%%%%%%%%%%%%%%%%%%%%%%%%%%%%%%%%
%%%%%%%%%%%%%%%%%%%%%%%%%%%%%%%%%%%%%%%%%%%%%%%

\section{Introduction}

The current morphology of moderately massive galaxies likely traces back to $z\sim 1-3$ \citep{conselice+05, buitrago+13} when the star formation rates of their progenitor galaxies were at their maximum (e.g., \citealt{leitner+11, yang+13, behroozi+13}).
By $z\sim 1$ the Hubble sequence was firmly in place  (e.g., \citealt{brinchmann+98, conselice+05, ilbert+06, oesch+10}) with ``peculiar'' galaxies, ubiquitous at higher redshift, being a sub-dominant galaxy population (e.g., \citealt{driver+95, scarlata+07, mortlock+13}).
The large diversity of galaxy morphologies encoded in the Hubble sequence, ranging from ellipticals to late-type spirals and irregulars, poses one of the key questions of galaxy formation and evolution:
Which mechanisms are responsible for the observed variety of galactic structural parameters, such as bulge-to-disc ratios, gas fractions, kinematics, and star formation rates (e.g. \citealt{hubble+26,sandage+61,dressler+80,devaucouleurs+91,roberts+94,steinmetz+02,gutierrez+04,driver+06,vanderwel+08,gavazzi+10,gavazzi+13})?

Galaxy interactions, such as minor and major galaxy mergers \citep{toomre+72, barnes+88, naab+03,boylankolchin+05,naab+06,cox+06,fiacconi+12,hilz+13}, galaxy harassment \citep{aguilar+85,moore+96,moore+98,mastropietro+05}, and tidal stirring of small galaxies by more massive ones \citep{mayer+01, lokas+10, kazantzidis+13}, are often suggested as important transformation channels.
In addition, secular processes induced by instabilities of the stellar component, such as bars, spiral arms or buckling/bending modes \citep{courteau+96,kormendy+04,debattista+06}, as well as gravitational instabilities of the gaseous disc \citep{dekel+09a,ceverino+10,bournaud+08,bournaud+11,bournaud+14,moody+14} may result in a transformation from late- to early-type morphology. 

Moreover, as the stellar component is largely built in-situ (e.g., \citealt{moster+13, behroozi+13}), galaxy morphology is strongly linked to the baryonic cycle of gas accretion, consumption, and ejection. 
Mechanisms of the latter category may regulate or even completely remove the gas content of a given galaxy, thus transforming the object from a blue, star-forming system to a red, passively-evolving one by quenching star formation.
Ram pressure stripping and strangulation \citep{gunn+72,abadi+99}, the transition from a cold to a hot mode accretion regime and gravitational quenching \citep{binboim+03,dekel+06, dekel+08}, outflows driven by stellar or AGN feedback \citep{scannapieco+05,bundy+08,dave+09,oppenheimer+10,teyssier+11,dubois+13}, and reduced gas accretion rates caused by cosmological starvation \citep{feldmann+14} likely all contribute given the right circumstances.

Simulations are of great help in understanding the qualitative and quantitative effect of such mechanisms.
Non-cosmological simulations starting from idealised initial conditions have spearheaded much of the progress thus far.
Part of the reason is that, until very recently, cosmological simulations were facing severe issues in realistically modelling galaxies of any sort \citep{mayer+08}.
The status of the field has recently improved, with different codes now producing massive spiral galaxies \citep{guedes+11,brook+11,agertz+11,stinson+13,agertz+13, hopkins+13a, agertz+14,marinacci+14}, dwarf galaxies \citep{governato+10,oh+11,shen+14, hopkins+13a} and even massive S0/ellipticals \citep{feldmann+10, oser+10, feldmann+11, oser+12, johansson+12,johansson+13,feldmann+14} with reasonably realistic properties.

These major improvements originate largely in the increased numerical resolution of zoom-in simulations combined with better sub-grid models of star formation and feedback processes (e.g. the Aquila and AGORA comparison projects; \citealt{scannapieco+12,kim+14}).
As a result, considerable progress has been made in matching a variety of predicted galaxy properties to observation (e.g. \citealt{mayer+12,bird+13,rashkov+13, hopkins+13a, agertz+14}). 
Furthermore, large scale cosmological simulations (e.g. \citealt{schaye+10, sales+12,mccarthy+12,vogelsberger+14,shy+14, schaye+14}) have recently been able to produce a range of galaxy morphologies and properties at $z\sim{}0$, an impressive result that was previously achieved only in zoom-in simulations for a small number of galaxies \citep{feldmann+11}. 
Unfortunately, the limited spatial resolution (typically $ \sim 1$ kpc) still hinders a robust quantitative analysis of internal structural and morphological evolution in large scale cosmological runs.

There are hints that significant morphological evolution, including bulge growth via bar instabilities triggered by minor mergers and tidal interactions, may start quite early, at $z>3$ \citep{guedes+13}.
However, works studying galaxy evolution at high-$z$ mostly concentrate on galaxies as massive as the Milky Way today at $z \sim 2-3$, which may undergo significant evolution via violent disc instabilities triggered by high gas accretion rates \citep{agertz+09,ceverino+12,moody+14}. 
These galaxies are the likely progenitors of some of the most massive present-day galaxies, such as those sitting at the centres of galaxy clusters and rich groups.

In this paper, instead, we extend the focus towards more typical galaxies with stellar masses between $10^8$ and $10^{10}$ M$_{\odot}$ at $z\sim 3-4$, which are likely progenitors of spiral galaxies and low- to intermediate-mass early-type galaxies today \citep{marchesini+09,ilbert+13,vandokkum+13}.
We primarily focus on the timescale and the mechanisms involved in generating their high redshift morphologies.
In particular, we ask when the Hubble sequence of such typical galaxies was first established and whether galaxy evolution at high $z$ is really different from low-$z$ as models of violent disc instabilities would suggest.

We address these questions using the Argo cosmological simulation, which fills the gap between zoom-in simulations of individual galaxies and large scale cosmological simulations since it has the resolution of the Eris simulation \citep{guedes+11} but a sub-volume almost ten times larger.
A predecessor simulation of Argo  carried out to $z=0$ at significantly lower resolution found clear evidence for many of the aforementioned mechanisms and identified a number of different evolutionary paths \citep{feldmann+11}.
However, the lower resolution limited the analysis to $z < 1.5$ because all galaxies except the central were poorly resolved before then. 

The Argo simulation follows a population of about two dozen $z\geq 3$ galaxies of intermediate mass in a representative region of the Universe.
Hence, the present work allows insights into the typical evolutionary paths of a (small) sample of such galaxies.
Galaxies of intermediate mass are ubiquitous at high-$z$ and future observatories such as the Atacama Large Millimeter/submillimeter Array (ALMA), the James Webb Space Telescope (JWST) and the European Extremely Large Telescope (E-ELT) should be able to probe the morphological evolution of these objects through high-resolution multi-band imaging and spectroscopy. 

The paper is organized as follows: in Section \ref{sec_2} we describe the main characteristics of the Argo simulations and the methodology used to identify and select halos and galaxies.
In Section \ref{sec_3} we present the results, focusing on the morphological evolution of our sample of galaxies, based on synthetic RGB images, circular velocity curves, surface density profiles and specific angular momentum analysis.
We highlight potential issues that may affect the conclusions drawn from our work in section \ref{sec_4}.
In Section \ref{sec_5} we discuss our main findings and conclude.

%%%%%%%%%%%%%%%%%%%%%%%%%%%%%%%%%%%%%%%%%%%%%%%
%%%%%%%%%%%%%%%%%%%%%%%%%%%%%%%%%%%%%%%%%%%%%%%
%%%%%%%%%%%%%%%%%%%%%%%%%%%%%%%%%%%%%%%%%%%%%%%

\section{Numerical simulations} \label{sec_2}

\subsection{Simulation code and initial conditions}

The Argo simulation is a suite of zoom-in cosmological simulations first presented by \citet{feldmann+14}.
The simulation is a follow-up of the $G2$ simulation described by \citet{feldmann+10} and \citet{feldmann+11}.
Argo has a high-resolution region of $\sim$3 comoving Mpc per edge centred on a dark matter halo of mass $\sim2 \times 10^{13}$~M$_{\odot}$ at $z=0$ inside a box of 123 comoving Mpc.
This dark matter halo hosts at $z=0$ a group dominated by a central, early-type, massive galaxy.
It is worth to point out that this specific group resides and evolves in a slightly over-dense but still rather usual environment since the matter over-density $\delta \equiv \rho/\langle \rho \rangle - 1$ (where $\langle \rho \rangle$ is the average matter density) at $z=0$ measured in $5 \, h^{-1}$~Mpc around the group is $\delta = 1.4$ and the final halo mass is smaller than the exponential cutoff mass in the halo mass function at $z=0$ and close to the halo mass ($\sim 10^{13}$ M$_{\odot}$) corresponding to 1-$\sigma$ fluctuations at $z=0$ (e.g. \citealt{reed+03, tinker+08, watson+13}).

The simulations were performed with the Tree/SPH $N$-body code GASOLINE \citep*{wadsley+04} assuming a \emph{Wilkinson Microwave Anisotropy Probe} 3-year cosmology (the same cosmology of the original $G2$ run) with $\Omega_{\rm m,0} = 0.24$, $\Omega_{\rm \Lambda,0} = 0.76$, $\Omega_{\rm b,0} \simeq 0.04$, $h= 0.73$, $\sigma_{8} = 0.76$ and $n = 0.96$ \citep{spergel+07}.
The suite consists of simulations at different resolutions and here we focus on the highest-resolution run (dubbed HR in \citealt{feldmann+14}).
  The force resolution is determined by the physical gravitational softenings $\epsilon_{\rm DM} = 250$~pc and $\epsilon_{\star} = 120$~pc for dark matter and baryonic particles, respectively.
The softenings are kept fixed in physical coordinates after $z = 9$.
The mass resolution of dark matter, gas and stellar particles is $m_{\rm DM} = 7.9 \times 10^5$~M$_{\odot}$, $m_{\rm g} = 2.1 \times 10^4$~M$_{\odot}$ and $m_{\star} = 6.3 \times 10^3$~M$_{\odot}$, respectively.
The initial conditions are made of 8,036,232 dark matter particles and 26,644,480 gas particles.
The simulation reaches redshift $z\simeq3.0$ with $\sim$54,000,000 particles.

GASOLINE follows the dynamical evolution of baryonic and dark matter due to gravity and hydrodynamics, taking into account the underlying expansion of the Universe.
It also implements sub-grid prescriptions for optically-thin radiative cooling, star formation, supernova feedback, mass loss from stellar winds and metal enrichment.
Since Argo HR has the same mass and force resolution, we employ the same parameters of the Eris simulation \citep{guedes+11}, a cosmological zoom-in simulation that successfully reproduced the properties of a Milky Way-like galaxy at $z=0$.
In particular, the gas is allowed to cool down to $\sim8000$~K following a radiative cooling function for an optically-thin gas with primordial composition \citep{wadsley+04}.
Note that we do not employ metal-line cooling in this work in order to be consistent with the code setup adopted for the original Eris simulation \citep{guedes+11,mayer+12}.
We included the effect of a uniform, redshift-dependent UV background \citep{haardt+96}.
The star formation is computed following \citet{stinson+06}: the local star formation rate is $\dot{\rho}_{\star} = \epsilon_{\rm SF} \rho_{\rm g} / t_{\rm dyn}$, where $t_{\rm dyn} = 1/\sqrt{G \rho_{\rm g}}$.
The gas is eligible to form stars if (i) it is in an over-dense, converging flow, (ii) it is locally Jeans-unstable,  (iii) the density $\rho_{\rm g} > \rho_{\rm th} = 5$~H~cm$^{-3}$ and (iv) the temperature $T_{\rm g} \leq 30000$~K.
The star formation efficiency $\epsilon_{\rm SF} = 0.05$.
Stellar particles are then stochastically spawned with initial mass $m_{\star}$ and represent a stellar population with a \citet{kroupa+93} initial mass function (also adopted by \citealt{kroupa+98}).
Stars add energy, mass and metals back to the surrounding gas via supernova feedback and stellar winds.
Both type Ia and type II supernovae inject $8 \times 10^{50}$ erg each to the neighbouring gas particles.
Type II supernovae originate from stars between 8 and 40 M$_{\odot}$ at the end of their life (determined from the parametrisation of \citealt*{raiteri+96}) and are modelled according to the analytical blast wave solution of \citet{mckee+77}.
In particular, the thermal energy is distributed to gas particles inside the maximum radius that a blast wave can locally reach and the cooling of those particles is shut off for the time corresponding to the end of the snowplow phase of the blast wave.
The cooling is not disabled for type Ia supernovae, the frequency of which is estimated from the binary fraction of \citet{raiteri+96}.
They eject all the same amount of mass (1.4 M$_{\odot}$) and metals (0.63 M$_{\odot}$ of iron and 0.13 M$_{\odot}$ of oxygen, \citealt{stinson+06}).
Low-mass stars between 1 and 8 M$_{\odot}$ return part of their own mass and metals to the surrounding gas through winds according to \citet{weidemann+87}.
The typical fraction of mass lost by a particle because of stellar winds is $\sim$40~\% (see \citealt{stinson+06} for additional details on the feedback model implementation).

%%%%%%%%%%%%%%%%%%%%%%%%%%%%%%%%%%%%%%%%%
% >>>>>> FIGURE : GROUP EVOLUTION <<<<<<
\begin{figure*}
\begin{center}
\includegraphics[width=18cm]{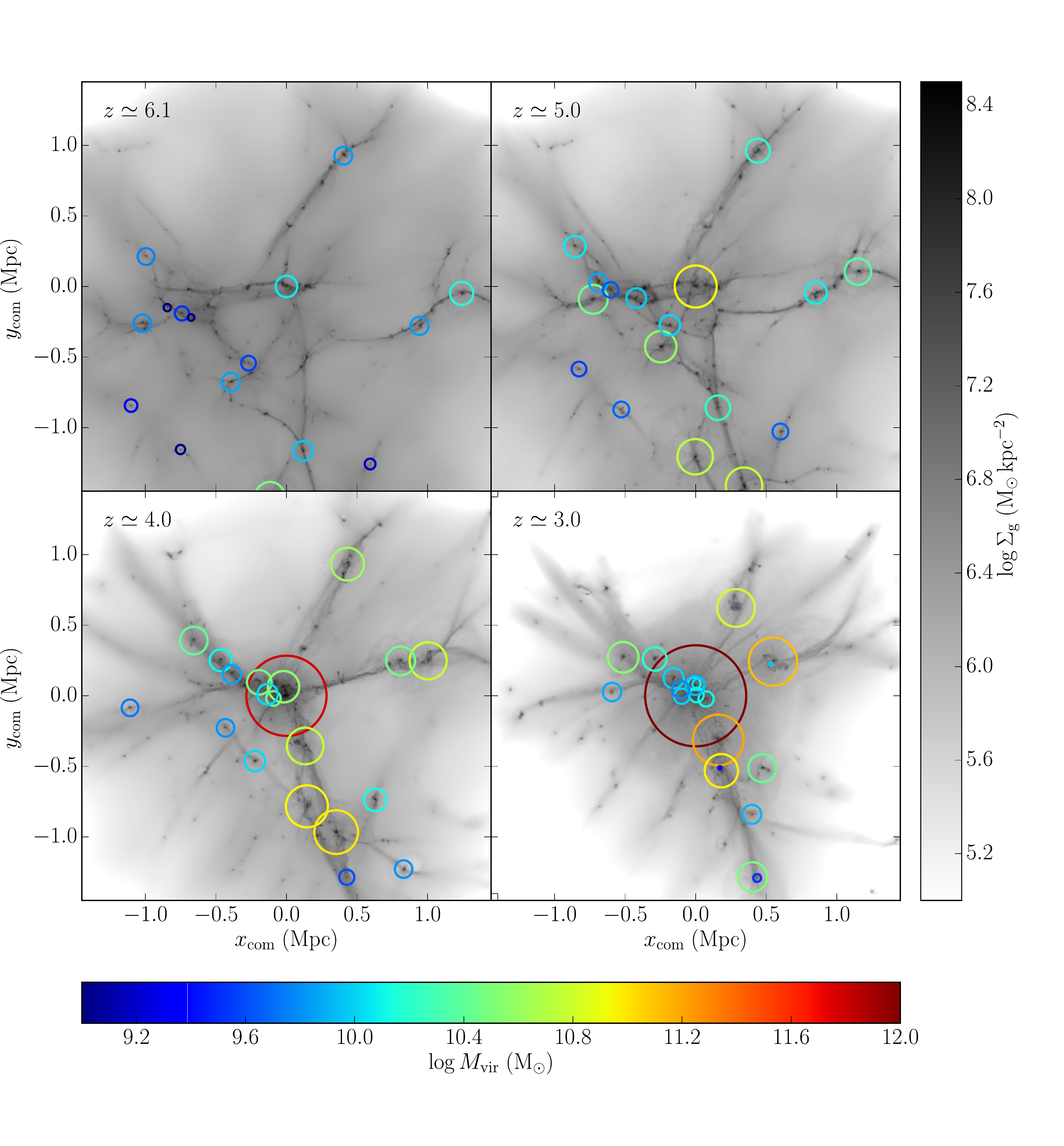}
\end{center}
\caption{Gas surface-density maps for the high resolution region of the Argo simulation. 
The panels show snapshots at $z\simeq 6$, 5, 4, and 3 (from top left to bottom right, see legend). 
Circles shows the positions of the 22 galaxies in our sample. 
The circle sizes and colours indicate virial radii and virial masses of the parent dark matter halos of the selected galaxies. 
All panels are centred on the progenitor of the primary halo, i.e., on the progenitor of the halo harbouring the most massive galaxy at $z=3$.}
\label{fig_group_evol}
\end{figure*}
% >>>>>>>>>>>>>>>>>>>>>> <<<<<<<<<<<<<<<<<<<<<<
%%%%%%%%%%%%%%%%%%%%%%%%%%%%%%%%%%%%%%%%%

%%%%%%%%%%%%%%%%%%%%%%%%%%%%%%%%%%%%%%%%%%%%%%%
%%%%%%%%%%%%%%%%%%%%%%%%%%%%%%%%%%%%%%%%%%%%%%%

\subsection{Halo detection and selection} \label{halo_detection}

We use the AMIGA Halo Finder (AHF, \citealt{gill+04, knollmann+09}) to identify dark matter halos and galaxies and to determine their properties, such as centres or virial radii.
The virial radius is defined as the radius enclosing a mean matter density $\Delta(z) \, \rho_{\rm crit}(z)$, where $\rho_{\rm crit}(z)$ is the critical density to have a flat Universe, whereas $\Delta(z)$ is the $z$-dependent virial over-density defined by \citet{bryan+98}.
We selected dark matter halos at $z\simeq3.4$ that (i) were more massive than $5\times{}10^{9}$ M$_{\odot}$ (corresponding to $>$6000 bound dark matter particles) and (ii) contained more than 10000 stellar particles.
Our selection results in a sample of 22 halos with virial masses between $5.8 \times 10^{9}$ and $1.1 \times 10^{12}$ M$_{\odot}$ hosting galaxies with stellar masses between $\sim 10^8$ M$_{\odot}$ and $\sim 10^{11}$ M$_{\odot}$.
We checked that these galaxies have at least $100$ star particles at $z \leq 10$ and typically $>$8000 star particles for $z \leq 4$.
Then, we traced them backward and forward in time, matching the dark matter particles with the same ID inside each halo for all snapshots between $z=10$ and $z=3$.
In particular, the main progenitor of an halo  in the snapshot $i$ is identified as the halo in the snapshot $i-1$ that maximise the value $f_{\rm shared} = N_{\rm shared} / \sqrt{N_{i} N_{i-1}}$, where $N_{i}$ and $N_{i-1}$ are the number of dark matter particles of the two halos in snapshot $i$ and $i-1$, respectively, and $N_{\rm shared}$ is the number of dark matter particles shared among the two halos.
Since dark matter particles have the same mass, $f_{\rm shared}$ can be interpreted as the average mass contributed by a halo to another halo.
Note also that $f_{\rm shared}$ and the procedure itself is independent on whether we proceed forward or backward in time with the matching.

%%%%%%%%%%%%%%%%%%%%%%%%%%%%%%%%%%%%%%%%%%%%%%%
%%%%%%%%%%%%%%%%%%%%%%%%%%%%%%%%%%%%%%%%%%%%%%%
%%%%%%%%%%%%%%%%%%%%%%%%%%%%%%%%%%%%%%%%%%%%%%%

\section{Results} \label{sec_3}

%%%%%%%%%%%%%%%%%%%%%%%%%%%%%%%%%%%%%%%%%%%%%%%
%%%%%%%%%%%%%%%%%%%%%%%%%%%%%%%%%%%%%%%%%%%%%%%

\subsection{Group evolution}

Figure \ref{fig_group_evol} shows the evolution of the high resolution region of the Argo simulation.
The main progenitor of the group-sized halo (the primary halo) forms at $z\gtrsim 7.5$ with a virial mass $\lesssim10^{10}$ M$_{\odot}$, comparable to the mass of a few other halos in the high-resolution region of the simulation at that redshift.
Between $z\sim{}6$ and $z\sim{}3.5$, the primary halo and its central galaxy grow quickly, outpacing the growth of other massive halos in the high-resolution region.
The primary halo lies at the crossing and merging point of several main dark-matter filaments that focus the dark matter and gas flows toward it, with a typical cold, star-forming gas ($T\leq30000$ K) inflow rate of $\sim 10$ M$_{\odot}$~yr$^{-1}$ through the virial radius at $z \sim 5.5$.
The cold gas penetrates inside $\sim 10\%$ of the virial radius, roughly where the central galaxy resides, sustaining an average star-formation rate of $\sim 20$ M$_{\odot}$ yr$^{-1}$ within it.
By $z\sim4$, the central halo is by far the biggest halo with a virial mass $\sim 4 \times 10^{11}$ M$_{\odot}$ and a physical virial radius $\sim 60$ kpc.
The main halo keeps growing up to $z \sim 3.5$ reaching $M_{\rm vir} \gtrsim 10^{12}$ M$_{\odot}$ due to increasing inflow rate of cold gas, but also due to accretion of small satellite galaxies that mostly form along the filaments and flow into the group, as shown in Figure \ref{fig_group_evol}.
After $z\sim3.5$, the main halo stops growing and enters the phase of ``cosmological starvation'' described by \citet{feldmann+14}.

Some of the other massive galaxies ($M_{\rm vir} \sim 10^{11}$ M$_{\odot}$ at $z\simeq3$) form even before the central one around $z\sim9-10$, but further away along the main filaments.
They move slowly along the filaments and smoothly accrete cold gas at a typical rate $\sim 5 -10$ M$_{\odot}$ yr$^{-1}$, similarly to the main halo, but almost steadily up to $z\sim3$. 
They typically experience up to 1-2 relevant merger episodes with mass ratios between $\sim$1:1 and $\sim$1:5 in the redshift range $3 \leq z \leq 10$.
Some of these galaxies also accrete a few gas-rich satellites with low stellar masses.

%%%%%%%%%%%%%%%%%%%%%%%%%%%%%%%%%%%%%%%%%%
% >>>>>> FIGURE : GALAXIES, FACE-ON <<<<<<
\begin{figure*}
\begin{center}
\includegraphics[width=18cm]{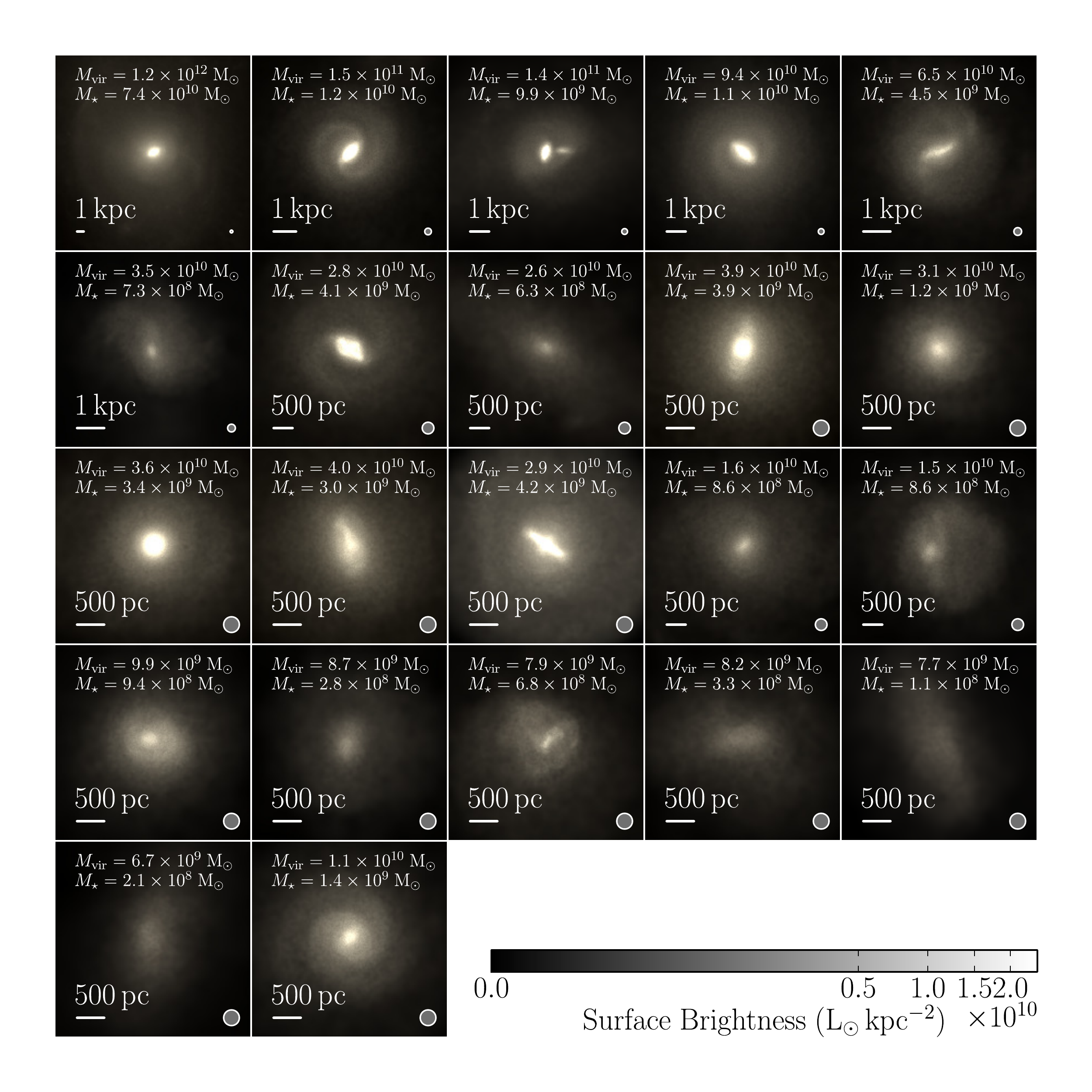}
\end{center}
\caption{
Face-on images of the 22 galaxies in our sample at $z\simeq3.4$ in the B, V and I filter bands.
Each panel encompasses roughly the inner 10\% of the virial radius of each halo. 
The bottom left corner shows the scale of each image in proper pc.
The stellar mass at $z \simeq 3$ and the virial mass at $z\simeq 3$ (or at the infall redshift for satellite galaxies at $z\simeq3$) are provided at the top of each panel.
The physical softening lengths of gas and star particles, $\epsilon_{\star} = 120$ pc, are indicated by the radius of the grey circle in the bottom right corner of each panel.
Many galaxies show spiral arms or bars within extended stellar discs.
Low mass galaxies often lack the concentrated central light concentration observed in more massive galaxies.}
\label{fig_gal_faceon}
\end{figure*}
% >>>>>>>>>>>>>>>>>>>>>> <<<<<<<<<<<<<<<<<
%%%%%%%%%%%%%%%%%%%%%%%%%%%%%%%%%%%%%%%%%%

%%%%%%%%%%%%%%%%%%%%%%%%%%%%%%%%%%%%%%%%%%
% >>>>>> FIGURE : GALAXIES, EDGE-ON <<<<<<
\begin{figure*}
\begin{center}
\includegraphics[width=18cm]{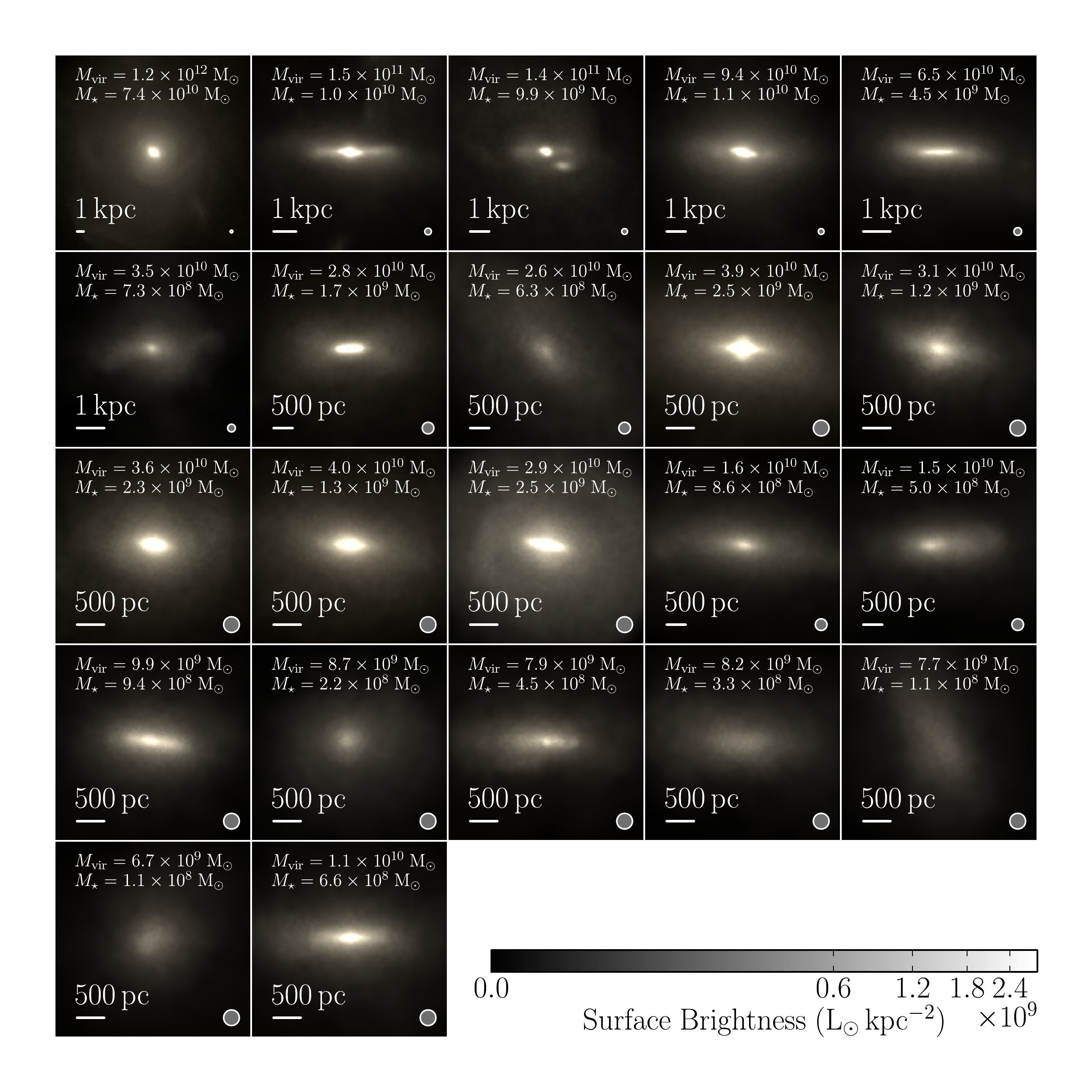}
\end{center}
\caption{
Same as Figure \ref{fig_gal_faceon}, but galaxies are shown edge-on.
Most galaxies (but not the central galaxy of the primary halo) have a pronounced disc component.
Lower mass galaxies have stellar discs that are thicker relative to the size of the galaxy.}
\label{fig_gal_edgeon}
\end{figure*}
% >>>>>>>>>>>>>>>>>>>>>> <<<<<<<<<<<<<<<<<
%%%%%%%%%%%%%%%%%%%%%%%%%%%%%%%%%%%%%%%%%%

%%%%%%%%%%%%%%%%%%%%%%%%%%%%%%%%%%%%%%%%%%%%%%%%%%%
% >>>>>> FIGURE : CIRCULAR VELOCITY PROFILES <<<<<<
\begin{figure*}
\begin{center}
\includegraphics[width=18cm]{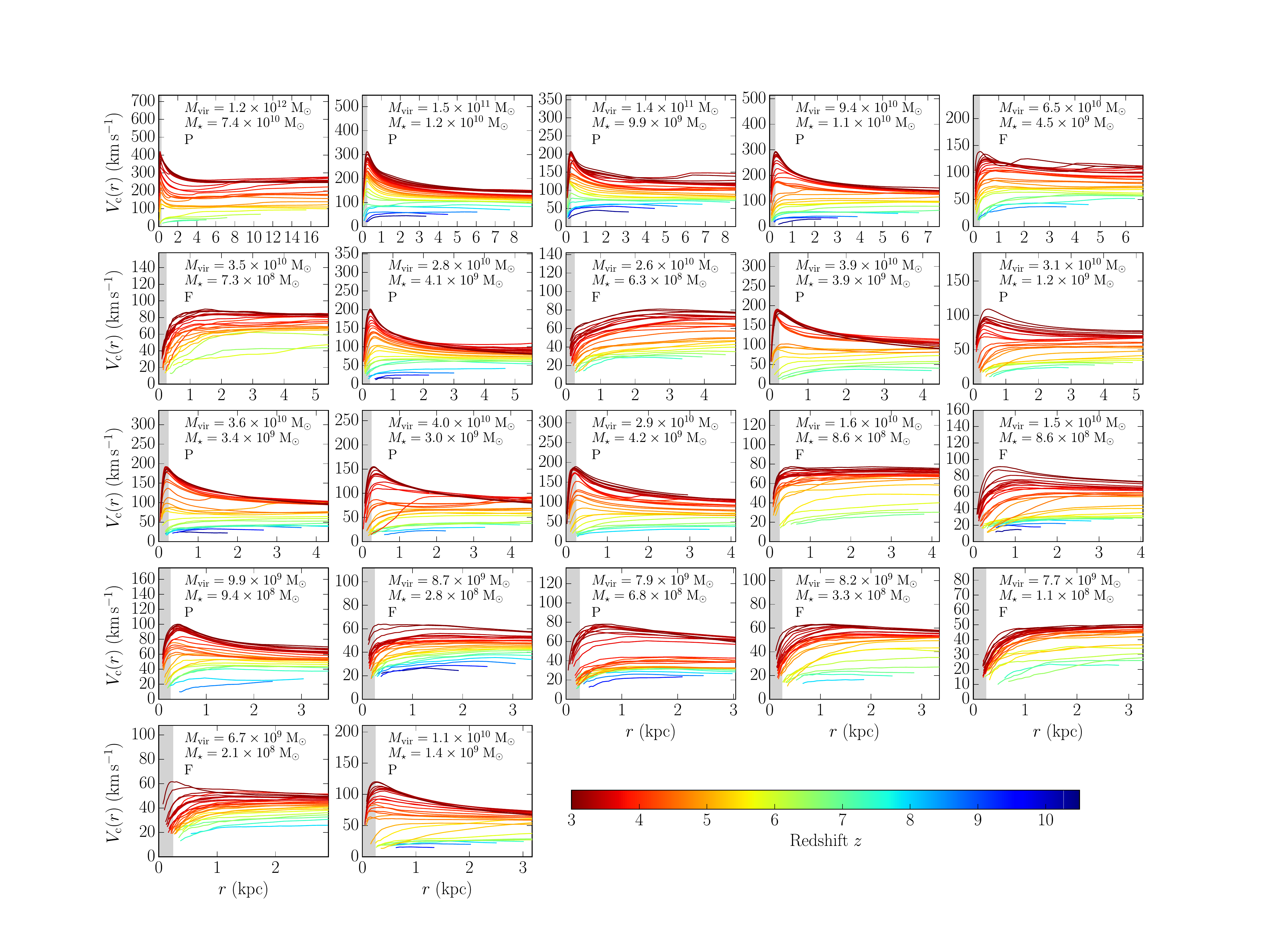}
\end{center}
\caption{
Circular velocity profiles for the 22 galaxies in our sample.
Curves in each panel are color-coded according to redshift (spanning the range $3 \leq z < 11$). 
The profiles are shown out to 20\% of maximally reached virial radius (radii are in physical units). 
The stellar mass at $z \simeq 3$ and the virial mass at $z\simeq 3$ (or at the infall redshift for satellite galaxies at $z\simeq3$) are provided at the top of each panel.
Each galaxy is flagged with ``F'' or ``P'' depending on whether it is classified as a ``flat'' or ``peaked'' galaxy at $z\sim{}3$.
The grey shaded areas correspond to 2 softening lengths, $r = 2 \epsilon_{\star}$.
Galaxies of lower (higher) stellar mass often show flat (peaked) circular velocity curves.}
\label{fig_vc_profile}
\end{figure*}
% >>>>>>>>>>>>>>>>>>>>>>>> <<<<<<<<<<<<<<<<<<<<<<<<
%%%%%%%%%%%%%%%%%%%%%%%%%%%%%%%%%%%%%%%%%%%%%%%%%%%

%%%%%%%%%%%%%%%%%%%%%%%%%%%%%%%%%%%%%%%%%%%%%%%
%%%%%%%%%%%%%%%%%%%%%%%%%%%%%%%%%%%%%%%%%%%%%%%

\subsection{Galaxy morphology} \label{sec_gal_morph}

Figure \ref{fig_gal_faceon} and \ref{fig_gal_edgeon} show face-on and edge-on mock images of our sample of galaxies at $z\simeq3.4$, respectively.
The face-on view is defined as the view along the line of sight determined by the specific angular momentum of stars inside $\sim 3\%$ of the virial radius $r_{\rm vir}$, excluding any contributions from satellites.
We assign a luminosity to each star particle from mass-to-light ratio tables\footnote{Tables available at \url{http://stev.oapd.inaf.it/cgi-bin/cmd}.} based on the isochrones and synthetic stellar population of the Padova group \citep{marigo+08, girardi+10, bressan+12}.
These tables span the stellar age interval from $4 \times 10^{6}$ to $12.6 \times 10^{9}$~yr, the metallicity interval from $5 \times 10^{-3}$ to $1.6$~$Z_{\odot}$ and assume the initial mass function from \citet{kroupa+93} and \citet{kroupa+98}.
We built the images using the computed emission in the I, V and B \citet{bessel+90} filters as the R, G and B channels, respectively, and we stretched the outcome with a power-law transformation (exponent $\gamma=0.3$) for better visualisation.

The central galaxy has an early-type, elliptical morphology, with thin stellar shells at a few physical kpc from the centre of the halo likely due to the tidal disruption of satellite galaxies on non-radial orbits \citep{hernquist+88, hernquist+89, feldmann+08}. 
The stellar mass at $z\sim3$ within $\sim10$ physical kpc is $\sim 7 \times 10^{10}$ M$_{\odot}$, with a total gas fraction (defined as the ratio between the gas mass and the sum of the stellar and gas mass) of $ \sim 11\%$. 
The fraction of cold ($\leq 30000$ K), star-forming gas is only $\sim 5\%$ and it is mainly distributed in a central disc with scale radius of $\sim450$ physical pc. 
See \citep{feldmann+14} for additional details on the central galaxy.

All the other galaxies exhibit a later-type morphology, ranging from extended, grand-design early-type spirals (likely Sa-Sb, from a visual inspection) to irregular, gas-dominated systems (Sm-Im, from a visual inspection).
All these galaxies have specific star-formation rates $\sim 2$ Gyr$^{-1}$ between $z \simeq4$ and 3, consistent with the observed star formation sequence at $z\simeq 3.7$ by \citet{lee+11}.
The most massive galaxies (virial masses $M_{\rm vir} \gtrsim$ a few $10^{10}$ M$_{\odot}$) typically have well-developed, grand-design spiral arms already at $z \gtrsim 3$.
Those spiral arms are likely triggered by nearby satellites.
However, numerical effects such as swing amplification induced by dark matter particles moving within the disc might also drive spiral structures.
\citet{donghia+13} have shown that swing-amplified, multi-armed spirals  can survive long after that the perturbation has faded away.
This effect might be present in low mass systems that show flocculent spiral arms.
The most massive disc galaxies often show a prominent bulge or a bar-like central distribution of stars, while lower mass galaxies are often pure discs.
The discs have typical scale heights of $h\sim400$ pc (a few softening lengths, estimated as the root mean square of the vertical displacement from the disc plane).
The discs are thin (aspect ratios $\ll{}1$) for the most massive and extended discs, while galaxies with $M_{\rm vir} \lesssim 10^{10}$ M$_{\odot}$ tend to develop thicker discs. 
This is likely a consequence of the stronger impact of stellar feedback in shallower potential wells. 
The stellar component of the lowest mass galaxies in our sample ($M_{\rm vir} < 10^{10}$ M$_{\odot}$, $M_{\star} \sim 10^{8}-10^{9}$ M$_{\odot}$) has a more irregular, although still flattened, morphology.

%%%%%%%%%%%%%%%%%%%%%%%%%%%%%%%%%%%%%%%%%%%%%
%%%%%%%%%%%%%%%%%%%%%%%%%%%%%%%%%%%%%%%%%%%%%

%%%%%%%%%%%%%%%%%%%%%%%%%%%%%%%%%%%%%%%%%%%%%%%%%%%%%%%%
% >>>>>> FIGURE : CIRCULAR VELOCITY RATIO VS MASS <<<<<<
\begin{figure*}
\begin{center}
\includegraphics[width=16cm]{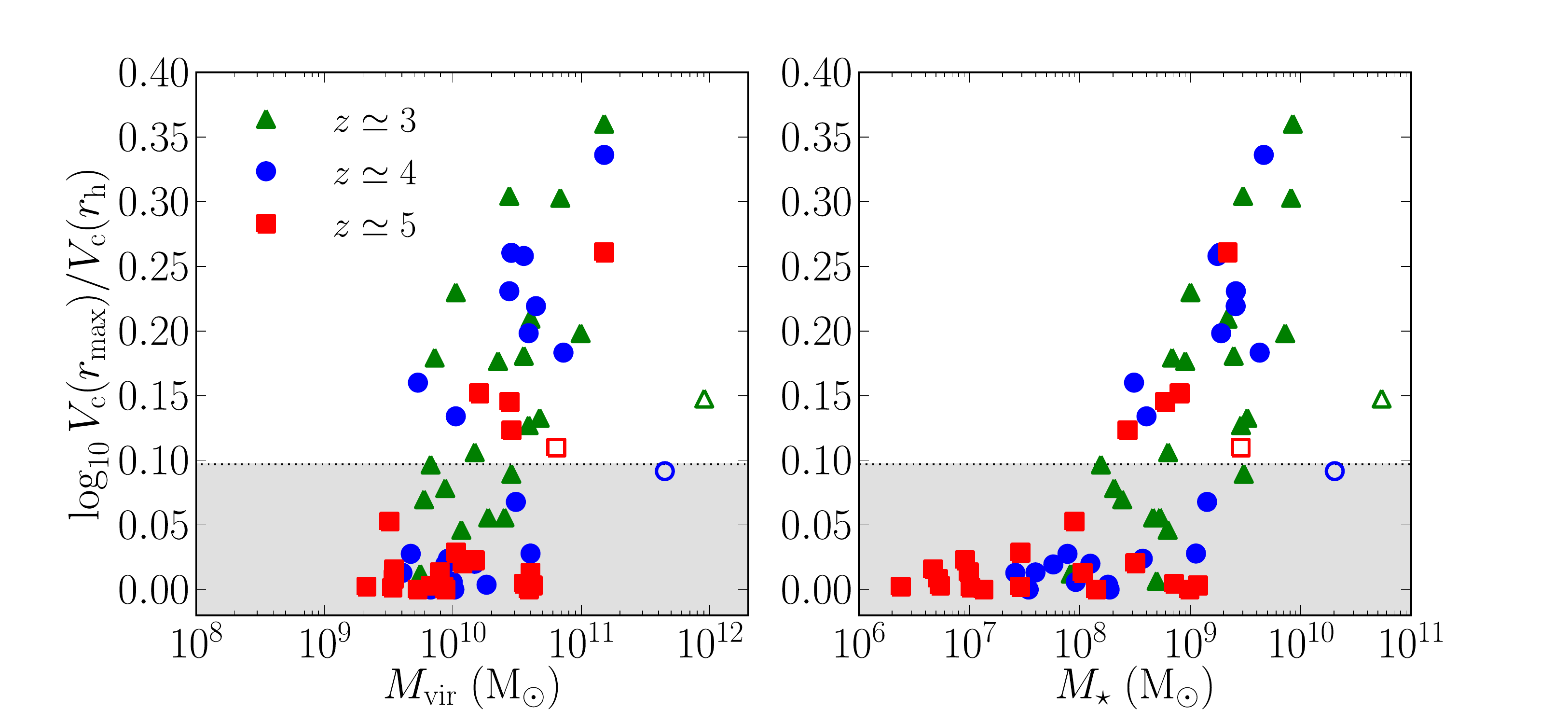}
\end{center}
\caption{Peakedness of the circular velocity profile as function of virial mass $M_{\rm vir}$ (left panel) and stellar mass $M_{\star}$ (right panel) for the 22 galaxies in our sample.
The peakedness is defined as $V_{\rm c}(r_{\rm max}) / V_{\rm c}(r_{\rm h})$.
Here, $r_{\rm max}$ is the radius at which the circular velocity peaks and $r_{\rm h}$ is the radius containing half the virial mass.
Red squares show the results for $z\simeq5$, blue circles for $z\simeq4$ and green triangles for $z\simeq3$.
The virial mass at infall time is used for satellite galaxies.
Empty symbols denotes the central galaxy of the primary halo.
The grey shaded area marks the region of ``flat'' systems, i.e., galaxies with $V_{\rm c}(r_{\rm max})/V_{\rm c}(r_{\rm h})<1.25$.
The peakedness of the circular velocity profile is correlated with the stellar mass of intermediate-mass galaxies and only weakly with their halo masses.}
\label{fig_vcratio_mass}
\end{figure*}
% >>>>>>>>>>>>>>>>>>>>>>>>>> <<<<<<<<<<<<<<<<<<<<<<<<<<<
%%%%%%%%%%%%%%%%%%%%%%%%%%%%%%%%%%%%%%%%%%%%%%%%%%%%%%%%

\subsection{Circular velocity curves}

Figure \ref{fig_vc_profile} shows the redshift evolution of the circular velocity profiles $V_{\rm c}(r) \equiv \sqrt{G M(r)/r}$, where $M(r)$ is the total mass inside the sphere of physical radius $r$, for the 22 galaxies in our sample.
The circular velocity profiles evolve quickly from $z\lesssim10$ to $z\simeq3$ via the rapid accretion of dark matter. 
For instance, the normalisation of the asymptotic velocity $V_{\rm c}(r_{\rm vir})$ increases from $\sim 20$ km s$^{-1}$ at $z=10$ to $\gtrsim 150$ km s$^{-1}$ at $z=3$ for galaxies with $z=3$ masses of $M_{\star} \sim{}10^{10}$ M$_\odot$.
At the same time, the shape of the circular velocity curves in the inner region is substantially affected by the distribution of baryons.
In particular, we can clearly distinguish two kinds of circular velocity profiles up to $z\simeq3$ in Figure \ref{fig_vc_profile}.
Some galaxies have flat or almost flat circular velocity curves (those galaxies will be dubbed as ``flat'' in the following), whereas other galaxies have a clear peak in the inner region (those galaxies will be dubbed as ``peaked' in the following).
In order to discriminate quantitatively between these two classes, we measured the ratio between the maximum circular velocity $V_{\rm c}(r_{\rm max})$ and the value $V_{\rm c}(r_{\rm h})$ at the radius $r_{\rm h}$ that encloses half of the \emph{total} mass of the halo.
We then classify a galaxy as ``flat''  or ``peaked'' at a given redshift if $V_{\rm c}(r_{\rm max})/V_{\rm c}(r_{\rm h})$ is below or above the assumed threshold of 1.25.
While the value 1.25 is to some degree arbitrary, our results are not strongly affected by the specific value that we assume.
Moreover, we decided to use $V_{\rm c}(r_{\rm h})$ instead of $V_{\rm c}(r_{\rm vir})$ because we found this definition more robust against the specific way the halo is identified and against the exact procedure to determine whether a particle is bound to a halo or not.

Figure \ref{fig_vc_profile} suggests that being a ``flat'' or ``peaked'' galaxy depends weakly on the virial mass of the host halo.
This is presented explicitly in the left panel of Figure \ref{fig_vcratio_mass}, which shows the ratio $V_{\rm c}(r_{\rm max})/V_{\rm c}(r_{\rm h})$ as a function of $M_{\rm vir}$ at three different redshifts.
We compute the Spearman's rank correlation coefficient to assess more quantitatively the degree of correlation between $V_{\rm c}(r_{\rm max})/V_{\rm c}(r_{\rm h})$ and $M_{\rm vir}$.
We find values between $\sim0.3$ and $\sim0.45$ from $z\simeq5$ to $z\simeq3$, implying that the correlation is weak.
This shows that having a ``flat'' or ``peaked'' circular velocity curve is neither strongly correlated with halo mass nor with the number of resolution elements (particles) per halo.
On the other hand, the right panel of Figure \ref{fig_vcratio_mass} shows the same plot with $M_{\star}$ instead of $M_{\rm vir}$.
In this case, the Spearman's coefficient ranges from $\sim0.4$ to $\sim0.8$ at the different redshifts, supporting the visual feeling that the ratio $V_{\rm c}(r_{\rm max})/V_{\rm c}(r_{\rm h})$ correlates more strongly with $M_{\star}$ than with $M_{\rm vir}$.

%%%%%%%%%%%%%%%%%%%%%%%%%%%%%%%%%%%%%%%%%%%%%%%%%%%%%%
% >>>>>> FIGURE : VCRATIO VS FORMATION REDSHIFT <<<<<<
\begin{figure}
\begin{center}
\includegraphics[width=8cm]{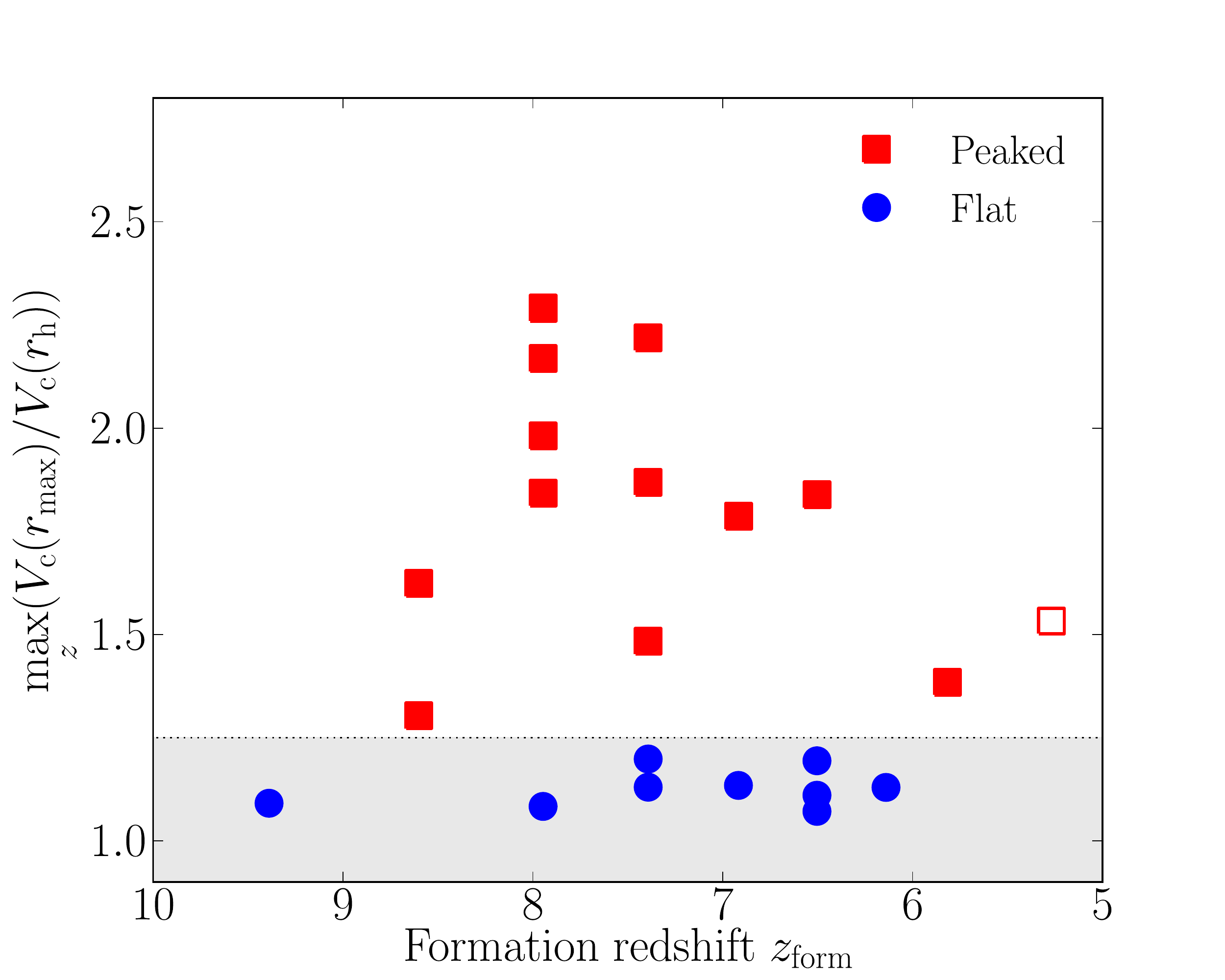}
\end{center}
\caption{
Maximum value of $V_{\rm c}(r_{\rm max})/V_{\rm c}(r_{\rm h})$ over the redshift range $3 \leq z < 11$ as a function of the formation redshift  $z_{\rm form}$.
Blue circles and red squares refer to ``flat'' and ``peaked'' galaxies, respectively.
The empty symbol denotes the central galaxy.
The grey shaded area marks the threshold $V_{\rm c}(r_{\rm max})/V_{\rm c}(r_{\rm h}) = 1.25$.
The peakedness of the circular velocity profile is not strongly correlated with formation redshift.}
\label{fig_vcratio_zform}
\end{figure}
% >>>>>>>>>>>>>>>>>>>>>>>>> <<<<<<<<<<<<<<<<<<<<<<<<<<
%%%%%%%%%%%%%%%%%%%%%%%%%%%%%%%%%%%%%%%%%%%%%%%%%%%%%%

%%%%%%%%%%%%%%%%%%%%%%%%%%%%%%%%%%%%%%%%%
% >>>>>>>> FIGURE : SD PROFILES <<<<<<<<<
\begin{figure*}
\begin{center}
\includegraphics[width=18cm]{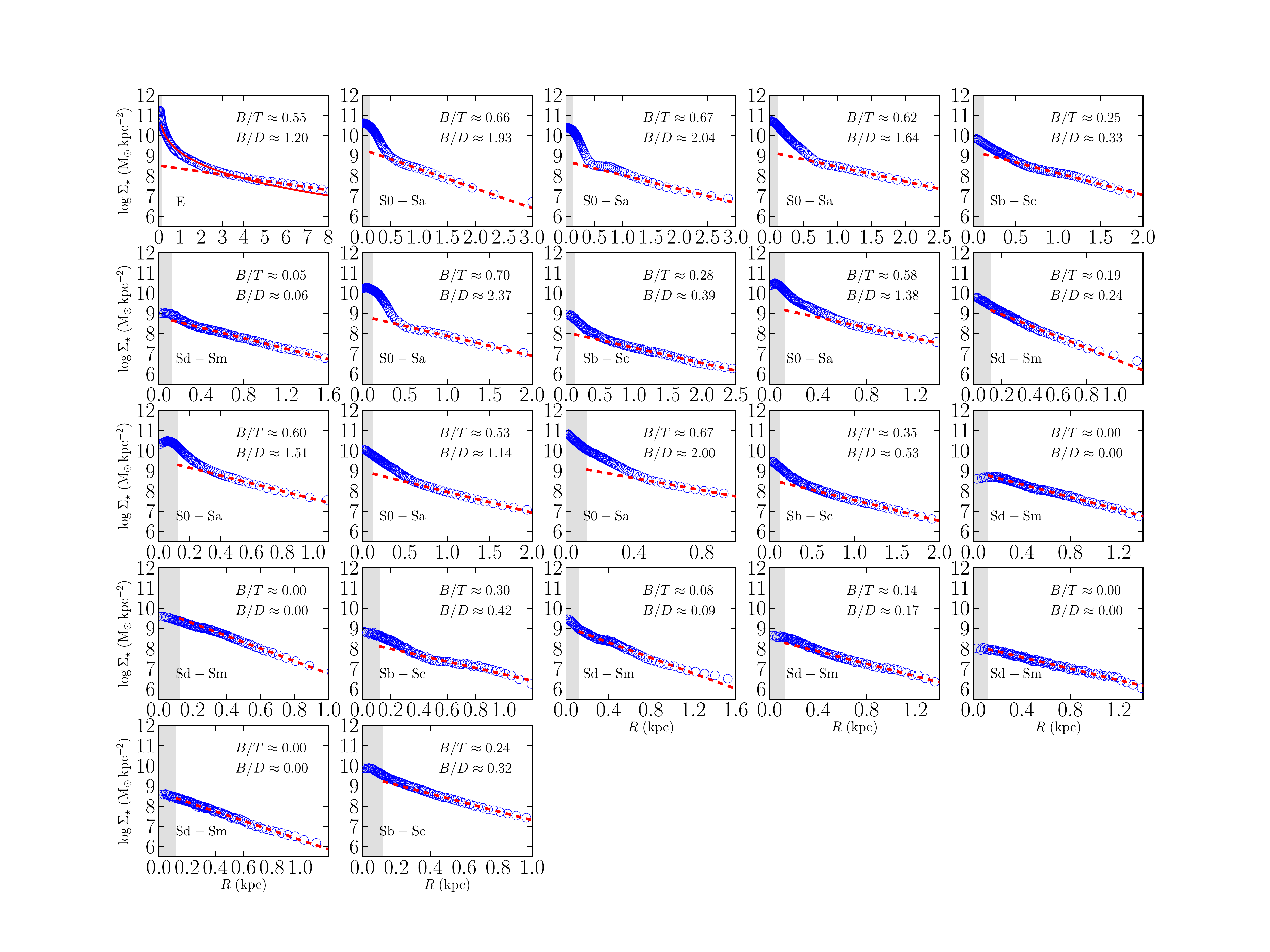}
\end{center}
\caption{
Stellar surface density profiles for the 22 galaxies in our sample at $z\simeq3$.
Blue circles show the data from the simulation, while red-dashed lines indicate the result of exponential fits to the stellar surface density at $r=0.05-0.1 r_{\rm vir}$.
The red continuous line in the first panel shows the de Vaucouleurs fit for the central galaxy used in Section \ref{sec_jstar}.
We show the inferred bulge-to-total ($B/T$) and bulge-to-disc ($B/D$) ratios at the top right of each panel.
We infer Hubble types based on the $B/T$ ratio and the presence of a stellar disc as described in the text.
The grey shaded areas correspond to a projected radius $\leq \epsilon_{\star}$.
Galaxies with higher stellar mass show a larger central excess, indicative of a larger $B/T$ ratio and an earlier Hubble type.}
\label{fig_SD_profile}
\end{figure*}
% >>>>>>>>>>>>>>>>>>> <<<<<<<<<<<<<<<<<<<
%%%%%%%%%%%%%%%%%%%%%%%%%%%%%%%%%%%%%%%%%

We can also test whether the shape of the circular velocity profile depends on the age of the galaxy.
In Figure \ref{fig_vcratio_zform} we plot the maximum over $z$ of the ratio $V_{\rm c}(r_{\rm max})/V_{\rm c}(r_{\rm h})$, the ``peakedness'' of the circular velocity profile, as a function of the formation redshift $z_{\rm form}$. 
The latter quantity is defined as the redshift at which the halo assembled at least $f_{\rm form}=5\%$ of its maximum mass at $z\geq{}3$.
The absence of a strong correlation between $z_{\rm form}$ and the peakedness of the circular velocity profile indicates that the shape of the mass distribution of our simulated galaxies is not merely related to the age of their parent halos.
In other words, a larger value of $z_{\rm form}$ does not necessary imply that a galaxy is more ``peaked''.
In particular,  we find both ``flat'' and ``peaked'' galaxies with any $z_{\rm form} > 6$.
Our results do not change qualitatively if we define $z_{\rm form}$ based on a different value for $f_{\rm form}$.
Our finding suggests that the evolution of the circular velocity profile is not simply dictated by secular evolution.

%%%%%%%%%%%%%%%%%%%%%%%%%%%%%%%%%%%%%%%%%%%%%
%%%%%%%%%%%%%%%%%%%%%%%%%%%%%%%%%%%%%%%%%%%%%

\subsection{The connection between $V_{\rm c}$ and morphology}\label{sec_morph}

%%%%%%%%%%%%%%%%%%%%%%%%%%%%%%%%%%%%%%%%%%%%%%%%%%%%
% >>>>>> FIGURE : B/T RATIO VS MASS & VCRATIO <<<<<<
\begin{figure*}
\begin{center}
\includegraphics[width=16cm]{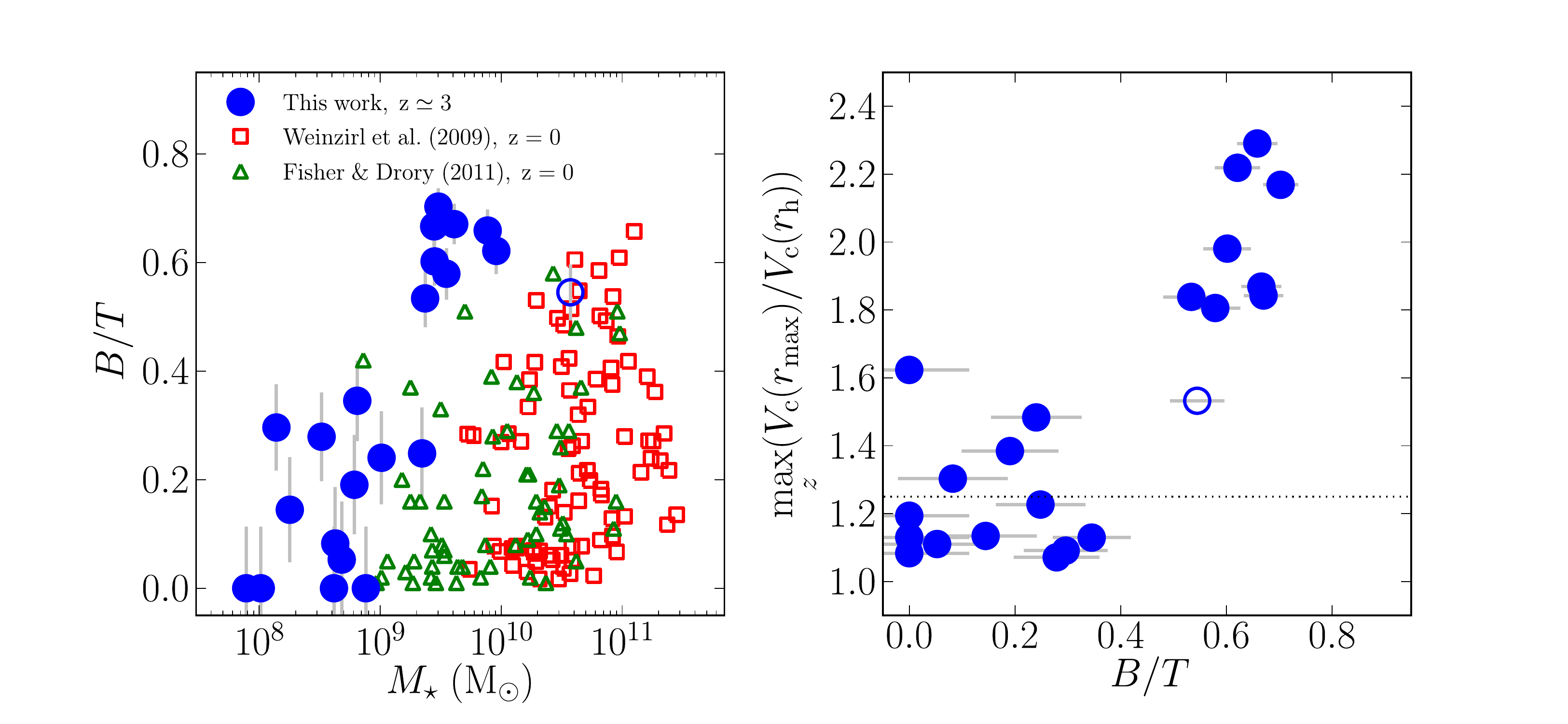}
\end{center}
\caption{$B/T$ ratio vs stellar mass at $z\simeq{}3$ (left panel) and  maximum peakedness of the circular velocity profile vs $B/T$ at $z\simeq{}3$ (right panel).
Blue circles show our galaxy sample, red squares are observed $B/T$ ratios by \citet{weinzirl+09} and green triangles are from \citet{fisher+11}.
The horizontal dotted line in the right panel marks the threshold $V_{\rm c}(r_{\rm max})/V_{\rm c}(r_{\rm h}) = 1.25$ that divides between ``flat'' and ``peaked'' galaxies.
We derive errorbars assuming 10\% uncertainty in the determination of both disc and total mass.
The empty circle denotes the central galaxy in each panel.
The peakedness of the circular velocity profile is a good indicator of the $B/T$ ratio of galaxies.
The morphology of the galaxies in our sample ranges from effectively bulge-less disc galaxies to spheroidal-dominated galaxies.}
\label{fig_BT_mass}
\end{figure*}
% >>>>>>>>>>>>>>>>>>>>>>>> <<<<<<<<<<<<<<<<<<<<<<<<<
%%%%%%%%%%%%%%%%%%%%%%%%%%%%%%%%%%%%%%%%%%%%%%%%%%%%

The circular velocity profile $V_{\rm c}(r)$ depends primarily on the mass distribution within $r$.
The mass distribution within 10\% of $r_{\rm vir}$ is dominated to a large extent by the baryonic (and in particular the stellar) mass.
Therefore, a connection between the central distribution of stars and the shape of the circular velocity curve is expected.
Figure \ref{fig_SD_profile} shows the stellar surface density profiles for the galaxies in our sample at $z\simeq3$.
The projection is face-on with respect to the axis given by the specific angular momentum of stars inside $0.03 r_{\rm vir}$.
We perform a ``partial'' profile decomposition based on the surface density profiles.
Instead of fitting the superposition of a bulge and a disc profile at the same time, we fit an exponential profile at large radii ($r\gtrsim 0.05 r_{\rm vir}$) to model a stellar disc, assuming that the bulge contribution to the surface density at these radii is negligible. 
We then infer the bulge mass by subtracting the mass of the fitted exponential profile from the overall stellar mass within $0.1 r_{\rm vir}$.
We choose this simplified procedure because many galaxies show complex nuclear morphology (e.g. stellar bars) that make difficult a full profile decomposition and we only aim at a single-parameter morphological characterisation.
Note that our determination of the bulge mass includes the mass of an eventual bar, as happens for many of the more massive galaxies of our sample (see Figure \ref{fig_gal_faceon}).
Moreover, our method bases the estimates of both the bulge and disc masses on the surface density profiles.
This might introduce some biases in the inferred $B/T$ and $B/D$ ratios with respect to what would be observationally determined via decomposition of surface brightness profiles.
In particular, we expect that our estimates are compatible with those obtained from near-infrared observations (e.g. in the H or K band), but they would likely overestimate the bulge component when compared with photometric measurements from bluer bands (e.g. B or V bands).
Finally, we do not attempt a kinematic decomposition between the bulge and the disc, which is also known to produce systematically different estimates of $B/D$ ratios \citep{scannapieco+10}.

Based on the values of $B/T$ (defined as the ratio of bulge mass to overall stellar mass), we assign to each galaxy a Hubble type: (i) E/S0-Sa for galaxies with $B/T \geq 0.4$, (ii) Sb-Sc for galaxies with $0.4>B/T\geq0.2$, and (iii) Sd-Sm for galaxies with $B/T < 0.2$. 
We distinguish between E and S0-Sa morphology based on the visual confirmation of an extended disc component.
Moreover, some low mass galaxies have $B/T \approx 0$, even though they might not appear as disc-dominated in Figure \ref{fig_gal_faceon} and \ref{fig_gal_edgeon}.
Nevertheless, they have a flattened stellar component and a high gas fraction distributed on a rotationally supported disc that qualify them as late-type, disc-like systems.
Therefore, although the face value $B/T \approx 0$ is a byproduct of our simplified procedure, this is still consistent with the typically observed $B/T\sim0.1-0.2$ for Sd and later-type galaxies (e.g. \citealt{weinzirl+09}) as long as we consider a reasonable uncertainty $\sigma \simeq 0.15$ on our $B/T$ estimates (we obtain $\sigma$ assuming 10\% uncertainty on both the bulge and disc masses, see below).
Although the exact $B/T$ values for the subdivisions are somewhat arbitrary, they are chosen in accordance with the morphological classification at $z=0$ \citep{scodeggio+02, graham+08, weinzirl+09}.

Comparing Figure \ref{fig_vc_profile} and \ref{fig_SD_profile} suggests that the most massive and ``peaked'' galaxies have higher values of the $B/T$ ratio.
We show this result more explicitly in Figure \ref{fig_BT_mass}, where we plot $B/T$ as function of stellar mass and of the peakedness of the circular velocity.
The errorbars on $B/T$ are derived assuming an uncertainty of 10\% for both the disc and total mass measurements.
The trend between $B/T$ and the stellar masses $M_{\star}$ is similar to the one found by \citet{weinzirl+09}, although their results are based on a sample of local spiral galaxies ($z=0$) of various morphological type (from S0 to Sm) and with larger stellar masses $\gtrsim10^{10}$ M$_{\odot}$ (determined from H band photometry).
We note that the $B/T$ values drawn from \citet{weinzirl+09} include the mass fraction of the reported bar component.
This is reasonable because we do not separate the non-disc component of our galaxies into a bar and a true bulge. We also compare our results with the data of \citet{fisher+11}.
Their sample at $z=0$ partially overlaps in mass with ours and shows a comparable range of $B/T$. 

As expected, the values of $B/T$ measured from the simulation are highly correlated with the maximum of $V_{\rm c}(r_{\rm max})/V_{\rm c}(r_{\rm h})$ over redshift. 
Hence, the evolution of the shape of the circular velocity curve is intimately connected with the morphological evolution, such that ``peaked'' galaxies are associated with more bulge-dominated galaxies. 
In other words, we can study the morphological evolution of galaxies by analysing the history of their circular velocity profiles.
Note, however, that we do not distinguish between pure bulges and pseudo-bulges as we do not perform a complete profile decomposition and we do not identify bulges based on their S\'{e}rsic indexes \citep{kormendy+04}.

%%%%%%%%%%%%%%%%%%%%%%%%%%%%%%%%%%%%%%%%%%%%%
%%%%%%%%%%%%%%%%%%%%%%%%%%%%%%%%%%%%%%%%%%%%%

\subsection{Morphology and the angular momentum content}\label{sec_jstar}

%%%%%%%%%%%%%%%%%%%%%%%%%%%%%%%%%%%%%%%%%
% >>>>>> FIGURE : J_star vs M_star <<<<<<
\begin{figure}
\begin{center}
\includegraphics[width=8cm]{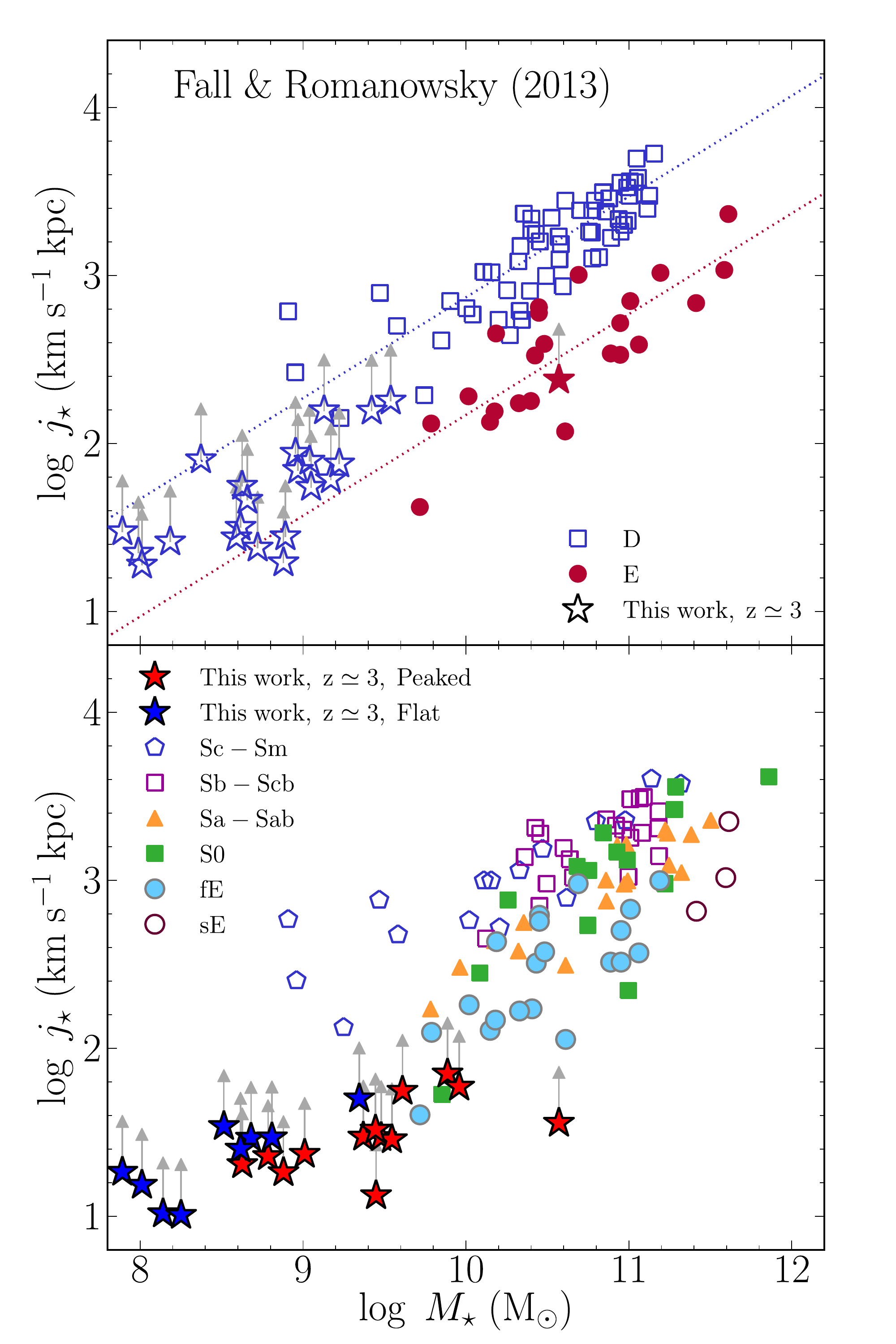}
\end{center}
\caption{
Stellar specific angular momentum $j_{\star}$ as a function of $M_{\star}$.
Upper panel: blue, empty symbols denote disc-only components (i.e. $j_{\star, \rm d}$ vs. $M_{\star, \rm d}$, see the text for details), whereas red, filled symbols denotes elliptical galaxies made by bulge only (i.e. $j_{\star, \rm b}$ vs. $M_{\star}$).
Empty squares and filled dots are observational data from \citet{fall+13}.
Stars are the data from our galaxy sample at $z\simeq 3$.
The upper and lower dotted lines represent the fits to the disc-only and bulge only relations, respectively (see \citealt{fall+13} for details).
Lower panel: stars denote our direct measure of the total $j_{\star}$ as a function of total stellar mass $M_{\star}$ (see the text for details).
Red stars mark ``peaked'' galaxies while blue stars mark ``flat'' galaxies.
The other symbols are observational data from \citet{fall+13} divided by morphological type according to the legend.
In both panel, grey vertical arrows show an increment of $j_{\star}$ by a factor of 2 at constant $M_{\star}$, consistent with the evolution with redshift predicted by Equation (\ref{eq_jstar_z}).}
\label{fig_jstar_mstar}
\end{figure}
% >>>>>>>>>>>>>>>>>>> <<<<<<<<<<<<<<<<<<<
%%%%%%%%%%%%%%%%%%%%%%%%%%%%%%%%%%%%%%%%%

\citet{romanowsky+12} and \citet{fall+13} recently revised the early findings of \citet{fall+83} that the stellar specific angular momentum $j_{\star}$ of galaxies correlates with the stellar mass $M_{\star}$ and that the correlation itself varies with morphological type.
We compare our simulation with these results in Figure \ref{fig_jstar_mstar}.
The upper panel of this figure reproduces Figure 2 of \citet{fall+13}, which shows the correlation between $j_{\star}$ and $M_{\star}$ for the disc component alone of (early- and late-type) spiral galaxies and for elliptical galaxies.
They estimate the disc-only contribution to $j_{\star}$, $j_{\star,\rm d}$, as $j_{\star,\rm d} = 2 V_{\rm rot, s} R_{\rm d} / \sin i$, where $R_{\rm d}$ is the scale radius of the exponential disc, $V_{\rm rot, s}$ is the asymptotic rotation velocity observed at large radii, and $\sin^{-1} i$ is a deprojection factor \citep{romanowsky+12}.
$j_{\star,\rm d}$ is plotted in Figure \ref{fig_jstar_mstar} against the disc mass $M_{\star, \rm d} = (1 - B/T) M_{\star}$.
They treat elliptical galaxies as ``pure bulges'' and then estimate $j_{\star}$ as $j_{\star, \rm b} = C k_{n} R_{\rm eff} V_{\rm rot, s}$, where $C$ is an inclination correction factor (1.21 for lenticulars and 1.65 for ellipticals), $k_{n}$ is a coefficient depending on the S\'{e}rsic index $n$, $R_{\rm eff}$ is the projected effective radius, and $V_{\rm rot, s}$ is the rotation curve evaluated at $\approx 2 R_{\rm eff}$ \citep{romanowsky+12}.

Mimicking the above procedure, we measure $j_{\star, \rm b}$ and $j_{\star, \rm d}$ for the central galaxy and the remaining galaxies of our sample, respectively.
First, we derive the gas rotation curves in the plane of the gaseous disc for all our galaxies so that we do not need any deprojection factor.
The gas rotation curve of the central galaxy is obtained in the plane of the central gaseous disc (see Section \ref{sec_gal_morph}).
In particular, we select a slice of $\sim 500$ pc centred in the plane of the disc (mimicking the action of a slit during the observation of an edge-on disc galaxy) and we measure the gas velocity as a function of radius projected along 20 random, polar line of sights (i.e. line of sights in the disc plane pointing toward the disc centre).
We finally average all the obtained rotation velocity curves\footnote{Since we measure the rotation velocity as the velocity projected on the line of sight in the plane of the disc, we do not need any deprojection factor because in our case $i=90^{\circ}$ and $\sin^{-1} i = 1$.}.
We then estimate $j_{\star, \rm d}$ of all but the central galaxy as $j_{\star, \rm d} = 2 R_{\rm d} V_{\rm s}$, where $V_{\rm s}$ is the asymptotic value of the gas rotation curve (typically similar, but slightly below, the asymptotic value of the circular velocity curve due to non-circular motions) and $R_{\rm d}$ is given by the exponential fits on the surface density profiles (Figure \ref{fig_SD_profile}).
Although we can decompose the surface density profile of the central galaxy in an exponential disc component and a remaining bulge component, a visual inspection clearly shows the lack of an extended disc.
Therefore, we treat this galaxy as an elliptical, ``pure bulge'' galaxy in the procedure described above.
We fit the surface density profile with a de Vaucouleurs profile (see Figure \ref{fig_SD_profile}) to determine $R_{\rm eff}$ and we compute $j_{\star, \rm b} = 2.29 V_{\rm s} R_{\rm eff}$, where $k_{4}=2.29$ and $V_{\rm s}$ is the rotation velocity at $\approx 2 R_{\rm eff}$.

We also estimate the total specific angular momentum following \citet{romanowsky+12} as $j_{\star} = (1-B/T) j_{\star, \rm d} + (B/T) j_{\star, \rm b}$, deriving the effective radii of bulges from the residual surface density profile after removing the exponential disc components.
Then, we compare these estimates with direct measurements of the stellar angular momentum of the galaxies, i.e. $j_{\star} = \sum \bmath{r} \times m_{\star} \bmath{v}$, where $\bmath{r}$ and $\bmath{v}$ are computed in the reference frame of the centre of mass of the galaxy.
We find values in fair agreement within a factor of $\sim 2$. 
We show our direct measurements in the lower panel of Figure \ref{fig_jstar_mstar}, which reproduces Figure 3 of \citet{fall+13}.
As expected, ``flat'' galaxies tend to have slightly higher $j_{\star}$ than ``peaked'' galaxies with comparable stellar mass.
This trend is similar to the trend with morphology revealed by the observational data.

%%%%%%%%%%%%%%%%%%%%%%%%%%%%%%%%%%%%%%%
% >>>>>> FIGURE : MERGER FIGURES <<<<<<
\begin{figure*}
\begin{minipage}{0.49\textwidth}
\centering
\includegraphics[width=9.5cm]{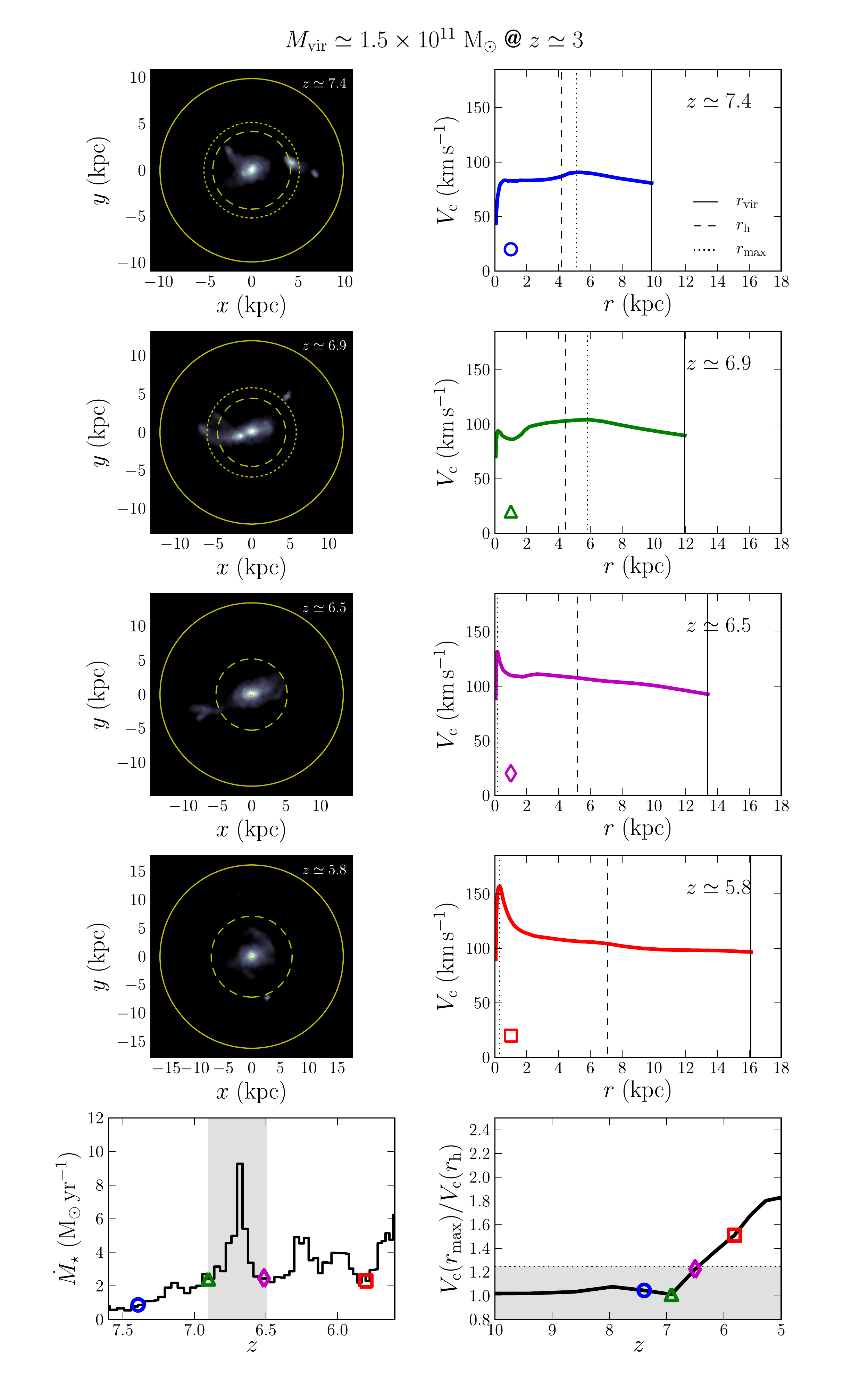}
\end{minipage}
\hfill
\begin{minipage}{0.49\textwidth}
\centering
\includegraphics[width=9.5cm]{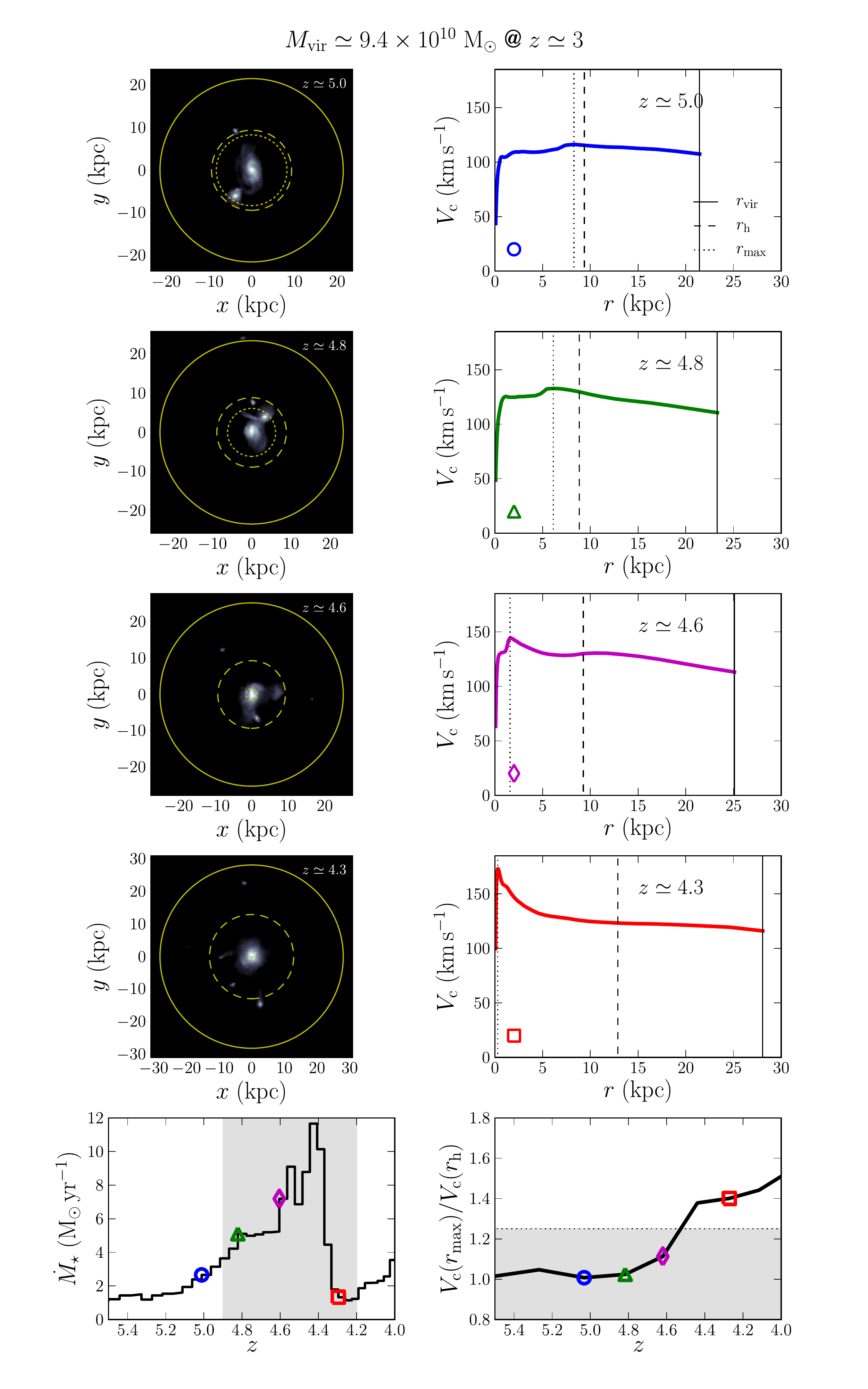}
\end{minipage}
\caption{
Evolution of the circular velocity curve for two example galaxies that obtain a peaked circular velocity profile via a major merger. 
The galaxy on the left (right) corresponds to the second (fourth) galaxy in the first row of Figure \ref{fig_vc_profile}.
For both example galaxies, we show the projected stellar density and the associated $V_{\rm c}(r)$ curve at four times during the galaxy interaction/merger (first four rows). 
Continuous, dashed and dotted lines mark the position of the virial radius, the position of the radius containing half of the total mass and the position of the radius corresponding to the peak of $V_{\rm c}$, respectively.
The last row shows the evolution of the star formation rate within the central kpc (first panel) and the evolution of $V_{\rm c}(r_{\rm max}) / V_{\rm c}(r_{\rm h})$ (second panel). In the bottom row panels blue circles, green triangles, magenta diamonds and red squares mark the redshifts of the corresponding four snapshots.
In both cases the peakedness of the circular velocity profile increases significantly over the course of the major merger.}
\label{fig_mergers}
\end{figure*}
% >>>>>>>>>>>>>>>>>> <<<<<<<<<<<<<<<<<<
%%%%%%%%%%%%%%%%%%%%%%%%%%%%%%%%%%%%%%%

Both the upper and the lower panel of Figure \ref{fig_jstar_mstar} show that our data sit on a relation similar to the observed one, with a slope $\alpha \sim 0.6$ for the disc-only components (upper panel).
However, the normalisation is different from that at $z=0$.
We argue that this discrepancy is due to the redshift evolution of this relation.
This can be qualitatively understood, at least for the disc-only components, using analytical models for galaxy formation.
In particular, following the simplest treatment of \citet{mo+98}, we can model a galaxy as an exponential, non-self-gravitating disc embedded in an isothermal halo, from which we derive the expected specific angular momentum:
\begin{equation}
j_{\star, \rm d} = \frac{(2 G)^{2/3} \lambda}{\Delta^{1/6}(z) H^{1/3}(z)} \left( \frac{f_{j}}{f_{\star}^{5/3}} \right) M_{\star}^{2/3},
\end{equation}
where $\lambda$ is the spin parameter of the host halo, $\Delta(z)$ is the $z$-dependent virial overdensity, $H(z)$ is the Hubble parameter, $f_{j}$ is the fraction of halo angular momentum retained by the disc, and $f_{\star}=M_{\star}/M_{\rm h}$ is the ratio between the disc and the halo mass.
We can then estimate the redshift evolution between $z=3$ and $z=0$ of $j_{\star, \rm d}$ at fixed $M_{\star}$ as:
\begin{equation} \label{eq_jstar_z}
\frac{ j_{\star, \rm d} (z=0) }{ j_{\star, \rm d} (z=3)} \Big|_{M_{\star}} = \left( \frac{\Delta (z=0)}{\Delta (z=3)} \right)^{1/6} \left( \frac{H (z=0)}{H (z=3)} \right)^{1/3} \sim 2.
\end{equation}
This calculation does not account for the presence of a bulge or for additional processes that might change $j_{\star}$ (e.g. mergers, feedback, etc.).
Nonetheless, it captures qualitatively the necessity of a redshift evolution for $M_{\star}-j_{\star}$ relation and is enough to explain and to cure the discrepancy between our $z\simeq 3$ data and the observations of nearby galaxies in the upper panel of Figure \ref{fig_jstar_mstar}.
The lower panel, instead, require a more detailed treatment of the build-up of galactic discs, although our simple correction is already enough to bring the simulated galaxies within the larger scatter expected at lower masses.
Note that $f_{j}$ and $f_{\star}$ can also contribute to the redshift evolution of the relation in an $M_{\star}$-dependent fashion.

%%%%%%%%%%%%%%%%%%%%%%%%%%%%%%%%%%%%%%%%%%%%%
%%%%%%%%%%%%%%%%%%%%%%%%%%%%%%%%%%%%%%%%%%%%%

\subsection{What triggers the difference between ``peaked'' and ``flat'' galaxies?}

Our sample of 22 galaxies at $z\geq3$ can be divided in two subclasses, ``flat'' and ``peaked'' galaxies, as discussed above.
Out of 22 galaxies, 9 never reach the threshold $V_{\rm c}(r_{\rm max})/V_{\rm c}(r_{\rm h}) = 1.25$ and they are thus classified as ``flat''.
The remaining 13 galaxies become ``peaked'' at redshift $z\geq{}3$, most of them at $z>4$.

All the galaxies that belong to the ``flat'' group, which represents $\sim40\%$ of our galaxy sample, have $V_{\rm c}(r_{\rm max})/V_{\rm c}(r_{\rm h}) \simeq 1$ almost all the time.
They are isolated galaxies that live in peripheral dark matter filaments.
They do not approach the central galaxy for more than 2 virial radii of the primary halo up to $z\simeq3$ and they do not experience any galaxy merger with mass ratios $\geq 1:5-1:6$.
Their star formation rates are slowly rising for $z>3$ and typically always below 1 M$_{\odot}$ yr$^{-1}$, apart from short, burst-like periods when they reach up to $\sim 2$ M$_{\odot}$ yr$^{-1}$.

On the other hand, the value of $V_{\rm c}(r_{\rm max}) / V_{\rm c}(r_{\rm h})$ for 8 out of 13 ``peaked'' galaxies, which represent $\sim60\%$ of the ``peaked'' subsample and $\sim36\%$ of our entire sample of galaxies, suddenly changes from $V_{\rm c}(r_{\rm max})/V_{\rm c}(r_{\rm h}) \simeq 1$ to $V_{\rm c}(r_{\rm max})/V_{\rm c}(r_{\rm h}) > 1.25$.
This change is clearly associated with a burst in star formation that can reach $\gtrsim 10$ M$_{\odot}$ yr$^{-1}$.
The starburst is triggered in each case by a major merger with a mass ratio $\geq 1:4$ for both $M_{\rm vir}$ and for $M_{\star}$, except for a couple of cases where the stellar mass ratio only is $<1:4$.

We demonstrate this finding in Figure \ref{fig_mergers}, where we follow the evolution of two example galaxies of our sample from before to after they merge with a lower mass companion galaxy.
In both examples, the peak circular velocity increases dramatically during the galaxy merger and the associated starbursting phase.
During that time the circular velocity profile transforms from flat to centrally peaked.
The radius at which the circular velocity reaches its maximum moves inwards from $r_{\rm h}\sim{}r_{\rm vir}/2$ ($\sim{}5-10$ kpc for the two example galaxies) to less than a kpc.  
These major mergers occur at different redshifts and in different environments.
The two galaxies in the first example have a total mass ratio $q\simeq1:2.9$ and merge around $z\simeq 6.7$ moving along a filament far from the primary halo.
The second example involves two galaxies with a total mass ratio $q\simeq1:3.6$. They merge around $z\sim 4.5$ in the vicinity (at $\sim 2 r_{\rm vir}$) of the primary halo, i.e., in a mildly overdense environment.

%%%%%%%%%%%%%%%%%%%%%%%%%%%%%%%%%%%%%
% >>>>>> FIGURE : BAR STRENGTH <<<<<<
\begin{figure}
\begin{center}
\includegraphics[width=8cm]{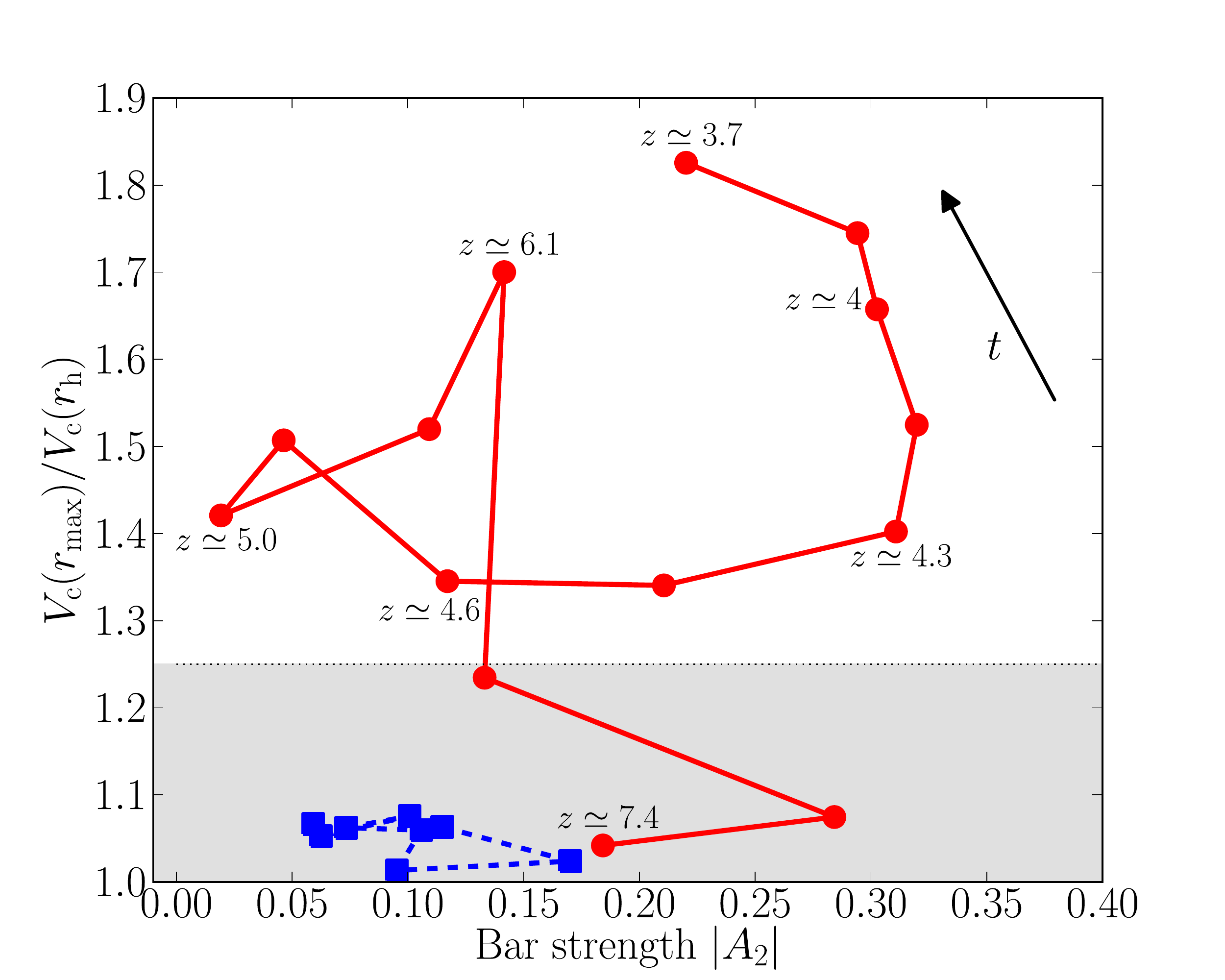}
\end{center}
\caption{Evolution of $V_{\rm c}(r_{\rm max})/V_{\rm c}(r_{\rm h})$, the peakedness of the circular velocity profile, as a function of the bar strength $|A_{2}|$.
The red solid curve corresponds to the third galaxy in the first row of Figure \ref{fig_vc_profile}.
This galaxy has a peaked rotation curve at $z\sim{}3.7$, but the growth of $V_{\rm c}(r_{\rm max})/V_{\rm c}(r_{\rm h})$ is not associated with a major merger event.
Instead, it gets peaked as a response to the formation of a stellar bar.
Reference redshifts are marked in the figure and the arrow indicates the direction of time along the curve.
The blue dashed curve shows the corresponding result for the fourth galaxy in the third row. 
This is a ``control'' galaxy with a flat circular velocity curve.
As expected no strong bar develops in this case.}
\label{fig_bar}
\end{figure}
% >>>>>>>>>>>>>>>>> <<<<<<<<<<<<<<<<<
%%%%%%%%%%%%%%%%%%%%%%%%%%%%%%%%%%%%%

The peaked circular velocity curves in the remaining five galaxies do not originate in major mergers.
These five cases represent $\sim38\%$ of the ``peaked'' subsample and $\sim22\%$ of the total sample.
The first of these anomalous galaxies is the central galaxy of the primary halo.
This galaxy undergoes a large number of repeated mergers (but not major mergers) and galaxy interactions with orbiting satellite galaxies of lower mass.
Its circular velocity profile starts to get strongly peaked only at relatively late times $z<4$. In fact, at $z\simeq4$ the circular velocity curve of this galaxy would be classified as ``flat'' ($V_{\rm c}(r_{\rm max}) / V_{\rm c}(r_{\rm h})\sim{}1.2$ at that time).
The second of the anomalous galaxies experiences a late ($z\sim{}3.6$) minor merger that is responsible for increasing $V_{\rm c}(r_{\rm max}) / V_{\rm c}(r_{\rm h})$ to just above 1.25.
The third anomalous galaxy enters the primary halo and obtains a peaked velocity profile during a starburst associated with its first pericentric passage.
In the remaining two anomalous galaxies we find that a bar drives the growth of a central stellar mass excess at high redshift ($z>5$).
At those times the bar is not properly resolved and, hence, the corresponding increase in $V_{\rm c}$ may be, at least in part, of numerical origin.
However, bars form also at later times (Figure \ref{fig_gal_faceon} clearly shows that many galaxies at $z\sim{}3$ are barred) and can drive the steady increase of the peakedness of the circular velocity profile.

As an example of the this process, we plot in Figure \ref{fig_bar} the evolution of the bar strength for one of the galaxies (third galaxy in the first row of Figure \ref{fig_gal_faceon}).
The bar strength is defined following \citet{dubinsky+09} as the absolute value:
\begin{equation}
\left| A_{2} \right| = \frac{1}{M} \left| \sum_{j=1}^{N} m_{j} \exp(2 i \phi_j) \right|,
\end{equation}
where the summation is performed over the $N$ star particles within a cut-off radius $R_{\rm c} = 500$ pc from the centre of the galaxy, $m_j$ and $\phi_{j}$ are, respectively, the mass and the polar angle in the plane of the galaxy of the $j$-th star particle, and $M = \sum_{j=1}^{N}m_j$.
The figure shows the evolution of $V_{\rm c}(r_{\rm max})/V_{\rm c}(r_{\rm h})$, the peakedness of the circular velocity profile, as a function of the bar strength $|A_{2}|$.
We find that the peakedness increases \emph{after} the bar forms, i.e., as a response to the presence of the bar. In turn, the bar gets weaker as the peakedness increases -- possibly because the formation of a bulge shuts down the passage of waves through the central region via the formation of an inner Lindblad resonance  which reduces the effectiveness of repeated swing amplification (e.g., \citealt{toomre+81, sellwood+89}).

%%%%%%%%%%%%%%%%%%%%%%%%%%%%%%%%%%%%%%%%%
% >>>>>> FIGURE : VCRATIO HISTOGRAM <<<<<<
\begin{figure}
\begin{center}
\includegraphics[width=8cm]{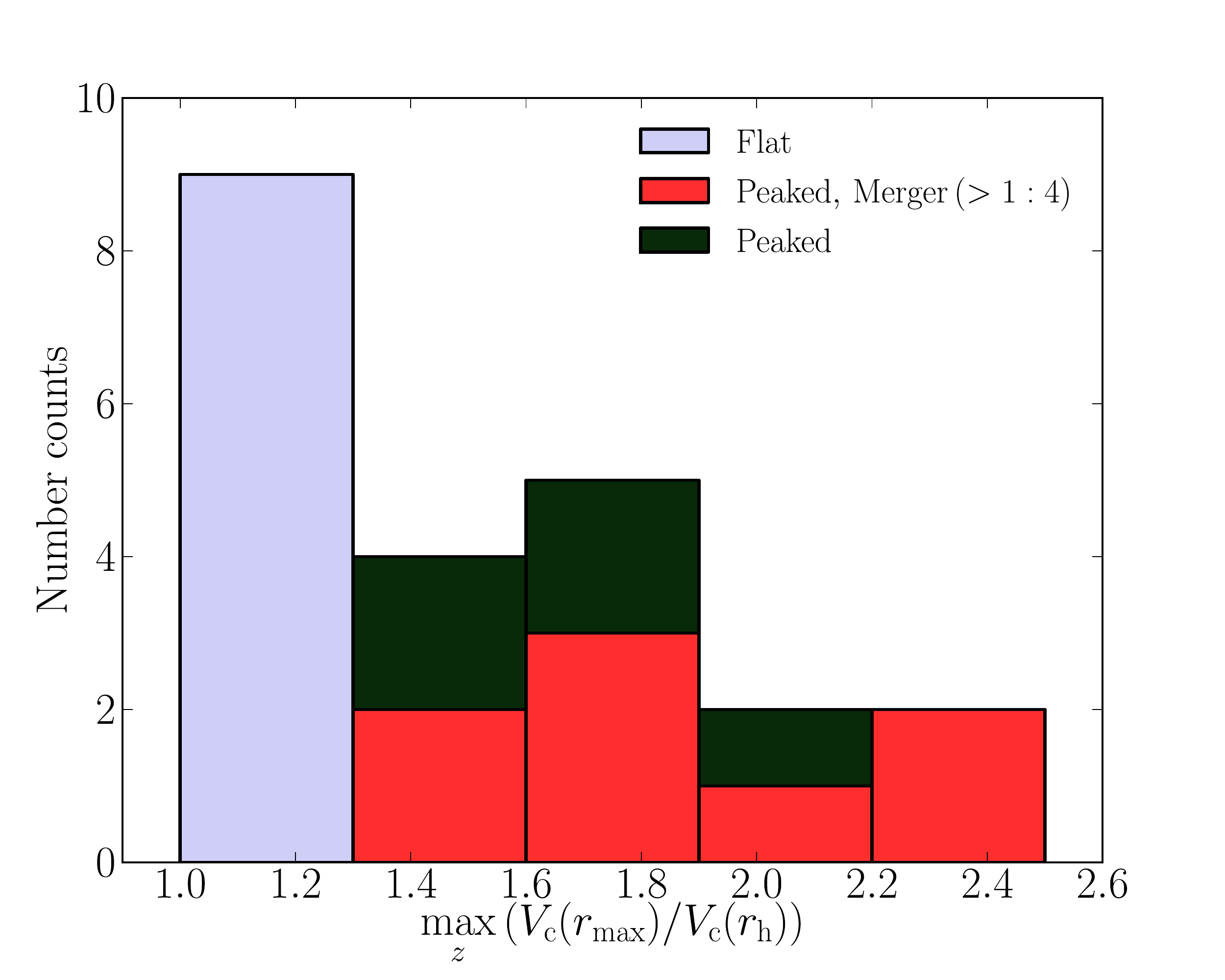}
\end{center}
\caption{Histograms of the distributions of the maximum $V_{\rm c}(r_{\rm max})/V_{\rm c}(r_{\rm h})$ over $z$.
Light blue bars show the number of galaxies with always flat rotation curves (9 out of 22).
Red bars indicate the distribution of galaxies that become peaked as a result of a major merger (8 out of 22).
Dark green bars show the distribution of galaxies that become peaked for other reasons (5 out of 22), see text.
Most galaxies that have a peaked circular velocity curve at late times undergo a major merger.
In contrast, none of the galaxies with a flat circular velocity profile undergo a major merger with mass ratio $\geq 1:4$.}
\label{fig_vcratio_stats}
\end{figure}
% >>>>>>>>>>>>>>>>>>>>>> <<<<<<<<<<<<<<<<<<<<<<
%%%%%%%%%%%%%%%%%%%%%%%%%%%%%%%%%%%%%%%%%

The discussion above is summarised by Figure \ref{fig_vcratio_stats}, which shows the distribution of $V_{\rm c}(r_{\rm max})/V_{\rm c}(r_{\rm h})$ for our galaxy sample.
Figure \ref{fig_vcratio_stats} suggests that major mergers might be the most effective process in shaping (and steepening) the central part of the circular velocity curve.
Given the connection between $V_{\rm c}$ and morphology discussed in Section \ref{sec_morph}, this in turn suggests that major mergers may be the primary process in assembling bulges with $B/T \geq 0.3$, in agreement with previous theoretical works (e.g. \citealt{naab+06, hopkins+10,kraljic+12}).
This is valid at least at high redshift ($z\geq3$), in slightly over-dense, group/proto-cluster environments and at mass scales ($M_{\star} \lesssim 10^{10}$ M$_{\odot}$) below the exponential cutoff mass of the stellar mass function at $z>2$ \citep{perez+08, marchesini+09, muzzin+13, tomczak+14}.

%%%%%%%%%%%%%%%%%%%%%%%%%%%%%%%%%%%%%%%%%%%%%%%
%%%%%%%%%%%%%%%%%%%%%%%%%%%%%%%%%%%%%%%%%%%%%%%
%%%%%%%%%%%%%%%%%%%%%%%%%%%%%%%%%%%%%%%%%%%%%%%

\section{Caveats} \label{sec_4}

In this study we probe the morphological evolution of a population of high-redshift galaxies formed in the Argo zoom-in cosmological simulation.
A key feature of this simulation is the high resolution and the big zoom-in sub-volume that allows us to follow the assembly of a normal galaxies population from redshift $\leq 10$ down to $z\simeq3$.
However, our simulation does not seem to be able to reproduce all the properties of the observed high redshift galaxies, as we discuss in the following of this section.
Nonetheless, we argue that our results are, at least in the qualitative picture, independent of and only mildly afflicted by these shortcomings.

We showed that the shape of the circular velocity curve correlates with the $B/T$ ratio and with stellar mass.
\citet{mccarthy+12} found similar results in terms of circular velocity curves' shape and trends with mass, but they attributed them to spurious overcooling effects above a certain stellar mass.
Although we can not exclude some residual overcooling, we are confident that this has only a  minor impact on our results, because (i) Argo has a much higher resolution with respect to the simulation presented in \citet{mccarthy+12}, in particular our softening is always smaller than the effective radius of the galaxies that we analyse, (ii) the most massive galaxy in our simulation, in which artificial overcooling should be most severe, matches nicely the predictions of abundance matching (see Figure \ref{fig_AM} below), (iii) we can attribute, in most cases, a physical cause to the sudden transformations of our galaxies, in particular to the increase of the central density and the associated steepening of the circular velocity curve.

%%%%%%%%%%%%%%%%%%%%%%%%%%%%%%%%%%%%%%%%%
% >>>>>> FIGURE : ABUNDANCE MATCHING <<<<<<
\begin{figure}
\begin{center}
\includegraphics[width=8cm]{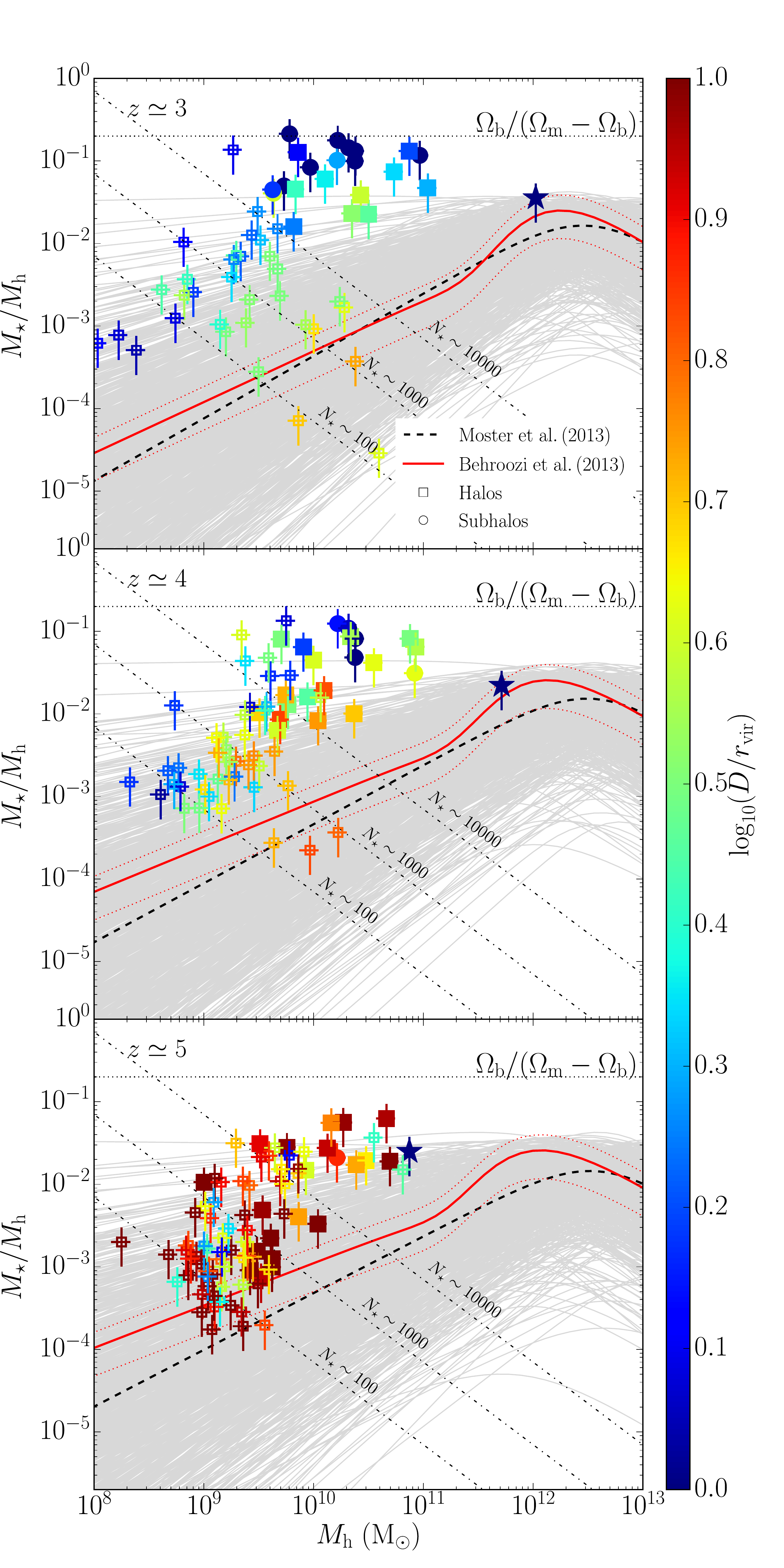}
\end{center}
\caption{Comparison with abundance matching predictions. 
We show the stellar mass -- halo mass relation at $z\simeq{}3$ (top), $z\simeq{}4$ (middle) and $z\simeq{}5$ (bottom).
Large filled symbols show the galaxies in our sample. 
Circles mark satellite galaxies and squares central galaxies.
Small empty symbols refer to central galaxies at each redshift with at least 100 star particles that are not included in our sample of 22 galaxies.
The star marks the central galaxy of the primary (group-sized) halo.
Symbols are color coded according to the distance of each galaxy from the primary halo in units of the virial radius $r_{\rm vir}$ of the primary halo.
Error bars assume 50\% and 20\% uncertainty on the $x$ and $y$ axis, respectively, following \citet{munshi+13}.
Black thick dashed line show the $z$-dependent stellar mass -- halo mass relation from \citet{moster+13} and the grey thin lines are 1000 Monte Carlo realisations of the same relation with the quoted errors.
The red continuous line is the $z$-dependent stellar mass -- halo mass relation from \citet{behroozi+13} and the red dotted lines show the 1-$\sigma$ deviation.
Dash-dotted lines show three different resolution limits for a given, approximate number of star particles.}
\label{fig_AM}
\end{figure}
% >>>>>>>>>>>>>>>>>>>>>> <<<<<<<<<<<<<<<<<<<<<<
%%%%%%%%%%%%%%%%%%%%%%%%%%%%%%%%%%%%%%%%%

A more serious challenge for our simulation is to reproduce the $M_{\star}-M_{\rm h}$ relation that is empirically determined with the help of the abundance matching technique \citep{behroozi+13,moster+13}.
In Figure \ref{fig_AM} we show the $M_{\star}-M_{\rm h}$ relation at $z\simeq 3$, 4 and 5 for the galaxies of our sample and for all central galaxies in the high resolution region of the Argo simulation that contain at least 100 star particles.
Stellar masses are measured within $0.1~r_{\rm vir}$ excluding any contributions from satellites.
$M_{\rm h}$ is typically the virial mass of the parent dark matter halo hosting the galaxy. However, for the satellite galaxies in our sample we use the halo masses at infall into their first host halo.

The figure shows that a large number (but not all) of the simulated galaxies lie above the empirically determined $M_{\star}-M_{\rm h}$ relation.
This discrepancy (i) grows with time, (ii) is larger for galaxies closer to the primary (group-sized) halo and (iii) is more pronounced for satellite galaxies than for central galaxies.
We therefore suggest that we are witnessing to some extent an environmental effect.
In particular, there are hints that the dark matter halo of galaxies in the vicinity of the primary halo might have undergone suppressed growth relative to the halo of a central galaxy of the same stellar mass, an effect ultimately exacerbated by progressive tidal truncation as galaxies enter the virial radius.
Further evidence supporting this interpretation is that the distance dependence of the discrepancy grows with time (see Figure \ref{fig_AM}).
If insufficient feedback were the single cause of the mismatch we would not expect to find a dependence on distance from the primary halo nor any time evolution of such a correlation.
Likewise, it is conceivable that there is an interplay between such effects and feedback, and therefore that the magnitude of such effects and thus the stellar mass assembled at a given time is still somewhat affected by the strength and mode of the feedback adopted.
Finally, as we mentioned above, the central galaxy, for which insufficient feedback should be especially problematic, nicely matches abundance matching predictions (see also \citealt{feldmann+14}).
We plan to study the effect of the environment on the $M_{\star}-M_{\rm h}$ relation in a forthcoming paper.

However, many of more massive galaxies at $z\simeq 5$ tend to be inconsistent with the abundance matching predictions regardless of the distance from the main halo.
It is also conceivable to expect that adding metal-line cooling would exacerbate the problem by increasing further the star formation efficiency, as suggested,
for example, by the somewhat higher mass of the spiral galaxy in the ErisMC simulation \citep{shen+12}, which included metal-line cooling, relative to the original Eris with only primordial cooling \citep{shen+12}.
While these galaxies have a somewhat biased formation history, given that they start within a slightly overdense Lagrangian region of the Universe and end up close to the primary halo at $z=3$, it is nonetheless plausible that the feedback model in our simulations is not efficient enough in low mass galaxies.
A conclusion along these lines has been suggested by \cite{stinson+13}, who argue that early stellar feedback (e.g., radiation pressure from massive stars) needs to be included in order to reduce the stellar fraction of high redshift galaxies.
The need for a different mode of feedback that delays star formation might be indicated also by the gas fractions found in the simulated galaxies.
These are in the range $\sim 5-70\%$.
In a number of cases, they fall below the typical values found for high-$z$ galaxies, which are in the range $20-80\%$ of the stellar mass (see e.g. \citealt{daddi+10,tacconi+10,debreuck+14}).
Nevertheless, we caution that measurements of gas fractions are usually available for galaxies significantly brighter than those in our sample, and they span a somewhat lower redshift, tipically below $z\sim 2.5$ \citep{daddi+10, tacconi+10, debreuck+14}.

Despite the aforementioned shortcomings, we maintain that the physical origin (major mergers) of the identified morphological transformations implies that our findings are robust to changes of the feedback physics. 
Nonetheless, it is likely that changes to the feedback model affect the timescales over which the morphologies of galaxies settle into their $z=0$ state.
It will clearly be helpful to compare our results with those of recent cosmological simulations that aim at modelling a variety of feedback processes at high resolution (e.g., \citealt{hopkins+13a}).

%%%%%%%%%%%%%%%%%%%%%%%%%%%%%%%%%%%%%%%%%%%%%%%
%%%%%%%%%%%%%%%%%%%%%%%%%%%%%%%%%%%%%%%%%%%%%%%
%%%%%%%%%%%%%%%%%%%%%%%%%%%%%%%%%%%%%%%%%%%%%%%

\section{Discussion and conclusions} \label{sec_5}

In this paper we investigated the properties of a sample of 22 high-$z$ galaxies during the assembly of a galaxy group in the Argo zoom-in cosmological simulation.
We focused on the morphological evolution of this galaxy sample as measured by a variety of (correlated) diagnostics such as the stellar surface-density profile, the $B/T$ ratio, the shape of the circular velocity profile and a visual morphological classification.
We found that the peakedness of the circular velocity profiles is a good proxy for the morphological classification.
In particular, galaxies with ``peaked'' circular velocity curves correspond to systems with more massive bulge components, while ``flat'' curves correspond to bulge-less disc galaxies of typically low stellar masses.
At the intermediate mass scale ($M_{\star} \lesssim 10^9$ M$_{\odot}$ and $M_{\rm vir}\sim{}0.5-5\times{}10^{10}$ M$_{\odot}$) a wide variety of rotation curves and $B/T$ ratios are present, showing that (halo) mass alone does not completely determine morphology.
Instead, the stellar mass is more closely correlated with morphology.

By analysing the origin of the dichotomy between ``flat'' and ``peaked'' galaxies we have identified major mergers as the main evolutionary trigger, with disc instabilities and minor accretion/interactions of satellites playing a sub-dominant role.
Note that this conclusion echoes the finding of \citet{feldmann+11} for the evolution of a galaxy population at much lower redshift ($z \sim 0-1.5$); indeed they identified mergers prior to infall inside the group potential as the main culprit behind the transmutation from discs into spheroids.

Our main conclusion is thus two-fold; (i) galaxy evolution in typical environments at high-$z$ appears to be surprisingly similar to galaxy evolution at low-$z$ despite the fact that galaxies are inherently more gas-rich and are fed by the cosmic web at higher gas accretion rates, and (ii) the Hubble Sequence is established very early during galaxy assembly, at $z \sim 3-4$, at least in the slightly biased regions around galaxy groups.

The central mass excess in our galaxies with peaked circular velocity profiles often appear to be exponential pseudobulges based on their stellar density profiles (although stellar bars can also contribute).
One notable exception is the central galaxy, which exhibits a profile that consistent with a de Vaucouleurs profile as expected for massive early-types, probably as a result of several repeated mergers/interactions \citep{naab+06,cox+06,moster+11}. 
This suggests that a mixture of dynamical and secular processes might be responsible for the formation of pseudobulges, supporting the findings of \citet{guedes+13}.

In the galaxies of our sample we do not observe the formation of giant star forming clumps via violent disc instabilities \citep{noguchi+99, dekel+09a, ceverino+10}. 
However, all but one of the galaxies in our sample are of low and intermediate mass, while clumpy galaxies typically belong to the more massive galaxy population.
Furthermore, our simulations do not allow gas cooling below $\sim 10^4$ K and the higher pressure hinders the gravitational collapse via a local Toomre instability \citep{ceverino+10}.
In addition, the effective stellar feedback in our simulation potentially suppresses the life times of giant clumps \citep{hopkins+12,hopkins+13b,moody+14}.
While galaxies with massive clumps are thus probably not the typical galaxies at high-$z$, a clumpy phase may still play an important role in the evolution of massive central galaxies in high density environments, e.g., for progenitors of present-day central group and cluster galaxies.

Our results suggest that the morphology of high redshift galaxies is determined by processes similar to those operating on/in local galaxies.
In particular major mergers, a natural consequence of the hierarchical structure formation in a $\Lambda$CDM Universe, appear to play a crucial role in the early morphological transformation of galaxies.
Future observations with ALMA, E-ELT, or JWST as well as larger samples of properly resolved, simulated galaxies may shed more light on the interplay between merging, feedback, and galaxy morphology as function of stellar mass and environment in high redshift galaxies.
Surveys with these new ground based and space born instruments will open new observational windows into the typical galaxy population at high redshift and be able to test our prediction that the Hubble Sequence near groups and clusters is already in place by $z = 2$.

%%%%%%%%%%%%%%%%%%%%%%%%%%%%%%%%%%%%%%%%%%%%%%%
%%%%%%%%%%%%%%%%%%%%%%%%%%%%%%%%%%%%%%%%%%%%%%%
%%%%%%%%%%%%%%%%%%%%%%%%%%%%%%%%%%%%%%%%%%%%%%%

\section*{Acknowledgements}

We thank the referee for the valuable comments that improved the paper.
We thank M. Fall and A. Rahmati for their careful reading of the draft and for their significant feedback.
We acknowledge the use of the Python package  pynbody (\url{https://github.com/pynbody/pynbody}) in our analysis for this paper.
We thank L. Girardi for maintaining his public website about stellar isochrones.
D.F. is supported by the Swiss National Science Foundation under grant \#No. 200021\_140645.
R.F. acknowledges support for this work by NASA through Hubble Fellowship grant HF-51304.01-A awarded by the Space Telescope Science Institute, which is operated by the Association of Universities for Research in Astronomy, Inc., for NASA, under contract NAS 5-26555.

%%%%%%%%%%%%%%%%%%%%%%%%%%%%%%%%%%%%%%%%%%%%%%%
%%%%%%%%%%%%%%%%%%%%%%%%%%%%%%%%%%%%%%%%%%%%%%%
%%%%%%%%%%%%%%%%%%%%%%%%%%%%%%%%%%%%%%%%%%%%%%%

\bibliographystyle{mn2e}
\bibliography{fiacconi_et_al}

%%%%%%%%%%%%%%%%%%%%%%%%%%%%%%%%%%%%%%%%%%%%%%%
%%%%%%%%%%%%%%%%%%%%%%%%%%%%%%%%%%%%%%%%%%%%%%%
%%%%%%%%%%%%%%%%%%%%%%%%%%%%%%%%%%%%%%%%%%%%%%%

\label{lastpage}

\end{document}